\documentclass{amsart}
\usepackage{amssymb, latexsym, color}

\definecolor{darkgreen}{rgb}{0,0.55,0}

\begin{document}

\newcommand{\chara}{{\mathbf{1}}}%{1\!\!1}
\newcommand{\barint}{\overline{\hspace{.65em}}\!\!\!\!\!\!\int}
\newcommand{\e}{\epsilon}
\newcommand{\ep}{\epsilon}
\newcommand{\logeps}{|\log\epsilon|}
\newcommand{\ue}{u_\epsilon}
\newcommand{\ve}{v_\epsilon}   
\newcommand{\jep}{j_\epsilon}
\newcommand{\loc}{ {\mbox{\scriptsize{loc}}} }
\newcommand{\R}{{\mathbb R}}
\newcommand{\C}{{\mathbb C}}
\newcommand{\T}{{\mathbb T}}
\newcommand{\Z}{{\mathbb Z}}
\newcommand{\N}{{\mathbb N}}
\newcommand{\calE}{{\mathcal{E}}}
\newcommand{\calC}{{\mathcal{C}}}
\newcommand{\calA}{{\mathcal{A}}}
\newcommand{\calD}{{\mathcal{D}}}
\newcommand{\calL}{{\mathcal{L}}}
\newcommand{\calF}{{\mathcal{F}}}
\newcommand{\calG}{{\mathcal{G}}}
\newcommand{\calH}{{\mathcal{H}}}
\newcommand{\calJ}{{\mathcal{J}}}
\newcommand{\calM}{{\mathcal{M}}}
\newcommand{\massnorm}{{\bf{M}}}
\newcommand{\flatnorm}{{\bf {F}}}
\newcommand{\dist}{\operatorname{dist}}
\newcommand{\sign}{\operatorname{sign}}
\newcommand{\extr}{\operatorname{extr}}
\newcommand{\supp}{\operatorname{supp}}
\newcommand{\rest}{{\, \bf{{  \llcorner}}\,}}

\newcommand{\beq}{\begin{equation}}
\newcommand{\eeq}{\end{equation}}

\theoremstyle{plain}
\newtheorem{theorem}{Theorem}
\newtheorem{proposition}[theorem]{Proposition}
\newtheorem{lemma}[theorem]{Lemma}
\newtheorem{corollary}[theorem]{Corollary}

\theoremstyle{definition}
\newtheorem{definition}{Definition}

\theoremstyle{remark}
\newtheorem{remark}{Remark}
\newtheorem{example}{Example}
\newtheorem{warning}{Warning}

\numberwithin{equation}{section}
\setcounter{tocdepth}{3}

%%% 
%%% end of newcommands etc
%%%

\author{S. Baldo \and  R.L. Jerrard \and G. Orlandi \and H.M. Soner}

\address{Department of Computer Science, University of Verona, Verona, Italy}\email{sisto.baldo@univr.it}
\address{Department of Mathematics, University of Toronto,
Toronto, Ontario, Canada}\email{rjerrard@math.toronto.edu}
\address{Department of Computer Science, University of Verona, Verona, Italy}\email{giandomenico.orlandi@univr.it}
\address{Department of Mathematics, ETH Z\"urich, Z\"urich, Switzerland}\email{mete.soner@math.ethz.ch}

\title[Vortex density models]{Vortex density models for superconductivity and superfluidity}

\date \today

\begin{abstract}
We study some functionals that describe the density of vortex lines in
superconductors subject to an applied magnetic field, and in Bose-Einstein
condensates subject to rotational forcing, in quite general domains in 3
dimensions. These functionals are derived from more basic models via
Gamma-convergence, here and in the companion paper \cite{BJOS1}. In our main results, we
use these functionals to obtain descriptions of the critical applied
magnetic field (for superconductors) and forcing (for Bose-Einstein),
above which ground states exhibit nontrivial vorticity, as well as a characterization of the vortex density in terms of a
non local vector-valued generalization of the classical obstacle problem.
\end{abstract}

\maketitle

\section{Introduction}

In this paper we study certain limits of  the Ginzburg-Landau model, which describes
a superconducting object in an external magnetic field, and the Gross-Pitaevsky functional,
which describes a Bose-Einstein condensate confined in a trap and subject to rotational forcing.

Most prior mathematical work on these sorts of problems has been limited to $2$-dimensional
models that are good descriptions, in various regimes, either of very flat, thin objects (superconductors or condensates), or of objects that are  translation-invariant, or very nearly so, in one direction.  Important results about the $2$-dimensional Ginzburg-Landau model,
obtained by Sandier and Serfaty in \cite{SandSerf2000a}, \cite{SandSerf2000c}
(see also the book \cite{SandSerfbook})
include the characterization of
the applied critical magnetic field, below which the ground state of a
superconductor expels the magnetic field, 
and above which the superconductor in the ground state is penetrated by magnetic vortices; and a description, in terms of an obstacle problem
solved by the magnetic field, of the limiting density of magnetic vortices 
above the critical applied field. Similar descriptions of 
2d Bose-Einstein condensates hold, though they are somewhat less well-documented in the literature.

In this paper we prove analogous results for the
full physical problems of $3$-dimensional superconductors and condensates.
In particular, we find a critical applied field for superconductors,
characterized by the same dichotomy as in the 2d case;
and we obtain a description, in the supercritical case, of the limiting vortex density in
terms of a constrained minimization problem solved by
the magnetic field. This problem is {\em not}
a classical obstacle  problem, but it is a kind of nonlocal, vector-valued obstacle 
problem with an interesting structure. %This problem formally reduces to the familiar 2d obstacle problem in the presence of suitable symmetries. 
We also establish corresponding
results about vortices in Bose-Einstein condensate wave functions, that is,
ground states of the Gross-Pitaevsky functional with rotational forcing.
These results include the determination of a critical rotational velocity,
and a characterization, in terms of a nonlocal generalization of an obstacle
problem, of the limiting vortex density for rotations
above this critical value.

We obtain these results from the study of  certain functionals whose ground states characterize vortex density, and other associated quantities, in
limits of sequence of minimizers of (suitably scaled) Ginzburg-Landau or
Gross-Pitaevsky functionals. In the case of Ginzburg-Landau, this limiting
functional was derived in a companion paper, see \cite{BJOS1}, as a corollary of a
general result proved there about the asymptotic behavior of a relatively simple model 
functional. In the case of Gross-Pitaevsky, the derivation of the limiting 
functional, using  results of \cite{BJOS1}, is given in Section \ref{S:becGlim}.

Although we mostly emphasize the analogy between the problems we study  here and 
obstacle problems, there are also close connections between our vortex density models and total variation models in image processing as introduced by Rudin, Osher, and Fatemi \cite{ROF}. (See \cite{CasChamNov} for a survey of related mathematical results.) 
In particular, the functional $\calG$ derived in Proposition \ref{prop:GPcompact}, see \eqref{G0.def}, can
be viewed as a generalization of the Rudin-Osher-Fatemi model, and
in situations with rotational symmetry, it reduces
to exactly a (weighted) Rudin-Osher-Fatemi functional. The paper concludes in
Section \ref{S:other} with a discussion of this and some related issues.

\subsection{the Ginzburg-Landau functional}

Let $\Omega$ be a bounded open subset of $\R^3$. A superconducting sample occupying the  region
$\Omega$ may be described by a pair $(u,A)$, where $u$ is a complex-valued function on $\Omega$ and 
$A$ is a $1$-form on $\R^3$,
that encode various physical attributes of the superconductor. For example,
$|u|^2$ corresponds to the density of Cooper pairs of superconducting electrons;
$d A$ can be identified with the magnetic field; and the superconducting current
is given by 
\beq
j_A u := \frac i2 ( u \overline{d_Au} - \bar u d_Au),\quad  \mbox{   where } \ d_A u := du - iAu.
\label{jAu.def}\eeq
Stable states of a superconductor in an external  magnetic field $H_{\e, ex} = d A_{\e,ex}$, with 
$A_{\e,ex} \in H^1_{loc}(\R^3; \Lambda^1\R^3)$,  correspond to minimizers (or local minimizers) of the
Ginzburg-Landau functional:
\beq\label{eq:Fep}
\calF_\ep(u,A)=\int_{\Omega}\frac{|d_A u|^2}{2}+\frac{(|u|^2-1)^2}{4\ep^2}\, dx+
\int_{\R^3}\frac{|dA-H_{\e, ex}|^2}{2}\, dx.
\eeq
Here the parameter $\e$ is related to physical properties of the superconducting sample. We will study the limit $\e\to 0$, with $A_{\e.ex}$ scaling so that it will turn out to be comparable to the critical value mentioned
above. For a discussion of the physical relevance of this scaling, see for example \cite{SandSerfbook}.

The model case is a constant external magnetic field, for which we may take $A_{\e, ex} =  \frac 12 c_\ep (x_1dx^2 - x_2 dx^1)$ for some real-valued
scaling factor $c_\ep$, corresponding to a spatially constant external field
$H_{\e,ex} = c_\e dx^1\wedge dx^2$, which in this example points in the $e_3$ direction.

The functional $\calF_\e$ makes sense for $u\in H^1(\Omega;\C)$ and $A$ such that
$A - A_{\e, ex}\in \dot H^1(\R^3;\Lambda^1\R^3)$. As is well known, the functional
is gauge-invariant in the sense that for any
such $(u,A)$ and for any function $\phi$ such that $d\phi \in \dot H^1(\R^3)$, the identity
$\calF_\e(u, A) = \calF_\e(e^{i\phi} u , A+ d\phi)$ holds. 
Moreover, $(u,A)$ and $(e^{i\phi}u, A+d\phi)$
correspond to {\em exactly} the same physical state, in
the sense that all physically observable quantities
are pointwise equal for the two pairs.

%\subsubsection{results: superconductivity}

\bigskip

Our starting point is the following, which is
an immediate consequence of
\cite{BJOS1}, Theorem 4. We use the notation
\[
\dot H^1_*(\Lambda^1 \R^3) := \{ \mbox{1-forms $A$ in } \dot H^1(\R^3) : d^*A = 0\}
\]
which is a Hilbert space with the inner product $(A, B)_{\dot H^1_*} := \int_{\R^3} dA\cdot dB$. We will often write $H^1_*$ for short. We  also write $H^k(\Lambda^p U)$ to denote the space of $p$-forms on $U$ with coefficients in the Sobolev space $H^k$.

\begin{proposition}\label{thm:3} 
Let $\Omega\subset\R^3$ be a bounded open set with $C^1$ boundary. Assume that $A_{\e, ex}\in H^1_{loc}(\Lambda^1R^3)$ and that there exists $A_{ex}\in H^1_{loc}(\Lambda^1\R^3)$
such that 
\beq\label{eq:Aeplim}
\frac {A_{\e, ex}} {\logeps} - A_{ex}\to 0 \quad\mbox{ in }\dot H^1_*(\Lambda^1\R^3)
:=\{ A\in \dot H^1(\Lambda^1\R^3) : d^*A=0\}.
\eeq
Let $(u_\e, A_\e)$ minimize $\calF_\e$ in $H^1(\Omega;\C)\times [A_{\e, ex} +  \dot H^1_*(\Lambda^1\R^3)]$.
Then there exists some $A_0\in  [A_{ex} +  \dot H^1_*]$ and $v_0\in L^2(\Omega; \Lambda^1\R^3)$ such that $dv_0$ is a measure, and such that 
\noindent
\begin{align}
\label{eq:convA}
&
\frac{A_\ep}{\logeps}- A_0\rightharpoonup 0 \quad\text{weakly in } \dot H^1_*(\Lambda^1\R^3),
\\
\label{eq:convju}
& \frac{j_{A_\ep}u_\ep}{\logeps }\rightharpoonup v_0 - A_0\quad\text{weakly in } {L^2(\Lambda^1\Omega)\, , }
\end{align}
and
\beq\label{eq:convJ}
\frac 1{\logeps}J_{A_\e}u_\ep := \frac 1{2 \logeps} d (j_{A_\e} u + A_\e) \to  \frac 12 dv_0\qquad\text{in }  W^{-1,p}(\Lambda^2\Omega)\, \quad\forall\, p\le 3/2 .
\eeq
\noindent
Moreover, $(v_0, A_0)$
minimizes the functional
 \beq\label{eq:F}
\calF(v,A)=\frac{1}{2}|dv|(\Omega)+\frac{1}{2}|| v-A ||^2_{L^2(\Lambda^1\Omega)}+\frac{1}{2}||dA-H_{ex}||^2_{L^2(\Lambda^2\R^3)}\, 
\eeq
in $L^2(\Omega;\Lambda^1\R^3)\times [ A_{ex} +  \dot H^1_*]$. Here $|dv|$ denotes the
total variation measure associated with $dv$. (We understand
$\calF(v,A)$ to equal $+\infty$ if $dv$ is not a measure.)
\end{proposition}

Our new results about superconductivity in this paper are derived
entirely by studying properties of the limiting functional $\calF$; the connection
to the more basic Ginzburg-Landau model  is provided by the above Proposition \ref{thm:3}.

Note that $\calF$ is gauge-invariant in the sense that if $\gamma\in \dot H^2(\Lambda^1\R^3)$, then 
\beq\label{Fgauge}
\calF(v - d\gamma|_\Omega, A+d\gamma) = \calF(v,A).
\eeq

Our first main result reformulates the problem of minimizing $\calF$ through convex duality,
the relevance of which in these settings was first pointed out in \cite{BS}.

\begin{theorem}
Let $\Omega\subset\R^3$ be a bounded open set with $C^1$ boundary, and assume that
$A_{ex}\in H^1_{loc}(\Lambda^1\R^3)$.
A pair $(v_0, A_0)$ minimizes $\calF$ in $L^2(\Lambda^1\Omega)\times [A_{ex}+ \dot H^1(\Lambda^1\R^3)]$ if and only if the following two conditions are satisfied:

\medskip\noindent
{\bf 1}. The $2$-form $B_0 = d(A_0 - A_{ex})$ belongs to the constraint set
\beq
\calC := \Big \{ B\in H^1(\Lambda^2\R^3)\cap d\dot H^1(\Lambda^1\R^3),\ \supp(d^* B )\subset\bar\Omega,
\ \ \| B\|_* \le \frac 12 \Big\},
\label{constraint}\eeq
where
\beq
\| B\|_* := \sup \Big \{ \int_{\R^3} B \cdot d\alpha \ : \ \alpha \in H^1(\Lambda^1\R^3), \int_\Omega |d\alpha| \le 1\Big\}.
\label{weirdnorm}\eeq
In addition, 
\beq
d^* B_0 + \chara_{\Omega} (A_0 - v_0) = 0 \qquad\mbox{ in }\R^3,%\qquad [B_N]=0 \mbox{ on }\partial\Omega,
\label{v0eqn}\eeq
%where $[B_N]$ denotes the jump of the normal component $B_N$ across $\partial\Omega$. 
%which implies in particular that 
and if $\bar v_0\in L^2(\Lambda^1\R^3)$ is any
$1$-form such that $\bar v_0|_\Omega = v_0$, then
\beq\label{eq:B0dv0}
\int_{\R^3} B_0\cdot d \bar v_0= - \frac{1}{2}\int_\Omega |d \bar v_0|.
\eeq

\medskip\noindent
{\bf 2}.  $B_0$ is the unique minimizer in $\calC$ of the functional
\beq
B \mapsto \calE_0(B) := \frac 12 \int_{\R^3}|B|^2 + \frac 12\int_{\Omega} |d^*B + A_{ex}|^2.
\label{dualfunctional}\eeq

\label{thm:GLmain}\end{theorem}

It is clear that if $\| B\|_*<\infty$ then
\beq
\int_{\R^3} B\cdot dv = 0\quad\quad\mbox{ for all $v\in H^1(\Lambda^1\R^3)$ such that 
$dv=0$ in $\Omega$}.
\label{Bbc}\eeq
Remark  also 
that \eqref{eq:B0dv0} implies that $\|B_0\|_*=\frac{1}{2}$  if the vorticity $dv_0\neq 0$. 
This is related to the  following necessary and sufficient condition for the vorticity to vanish.

\begin{theorem}Let $\Omega\subset\R^3$ be a bounded open set with $C^1$ boundary, and assume that
$A_{ex}\in H^1_{loc}(\Lambda^1\R^3)$.
Let $(v_0, A_0)$ minimize $\calF$ in $L^2(\Lambda^1\Omega)\times [A_{ex}+ \dot H^1(\Lambda^1\R^3)]$, and let $B_*$ denote the unique minimizer of
%Then $dv_0 = 0$ if and only if the unique minimizer $B_*$ of
$\calE_0(\cdot)$ in the set 
\beq
\calC ':=\Big \{ B\in H^1(\Lambda^2\R^3)\cap d\dot H^1(\Lambda^1\R^3) : 
B\mbox{ satisfies }\eqref{Bbc} \}%\supp(d^* B )\subset\bar\Omega \Big\},
\label{constraint'}\eeq
where $\calE_0$ is defined in \eqref{dualfunctional}.

Then $dv_0=0$ if and only if $\| B_*\|_* \le \frac 12$.

As a result, if  $(u_\e, A_\e)$ minimizes $\calF_\e$ in $H^1(\Omega;\C)\times [A_{\e, ex} +  \dot H^1_*(\Lambda^1\R^3)]$, then $\logeps^{-1}J_{A_\e}u_\e\to 0$ as $\e\to 0$ in $W^{-1,p}, p<3/2$,
if and only if $\|B_*\|_* \le \frac 12$.
\label{cor:critfield}\end{theorem}

The subtle point in Theorem \ref{cor:critfield} is the identification of the correct space 
$\calC'$ in which the magnetic field for the vortex-free minimizer is energetically optimal.

\begin{remark}
We  give a different (but necessarily equivalent) characterization of when $v_0$ is vortex-free,
and a different dual problem, see Theorem \ref{thm:GLcritical} and Lemma \ref{L.dualb} in Section \ref{S:3.2}.  
\label{R1}\end{remark}

\begin{remark}
Observe that for $B\in\calC'$ we have, by virtue of Hahn-Banach theorem, 
$$\|B\|_*=\inf\{ \|\beta\|_{L^\infty(\Omega)},\ \ \beta\in H^1(\Lambda^2\R^3),\ \ d^*\beta=d^*B\},$$
so that $\|B\|_*$ can be interpreted as a nonlocal $L^\infty$ norm of $B\in \calC'$ for $\Omega\subset\R^N$ with $N\ge 3$,
while for $\Omega\subset\R^2$ we have $\|B\|_*=\|B\|_{L^\infty(\Omega)}$, since in that case in \eqref{weirdnorm} one can test with forms $\alpha_j\in H^1(\Lambda^1\R^2)$ with $d\alpha_j=f_j(x)dx_1\wedge dx_2$ such that the scalar functions $f_j(x)$ converge to  a Dirac mass $\delta_{x_0}$ for arbitrarily fixed $x_0\in\Omega$.

Hence in the 2-dimensional case $\Omega\subset\R^2$ the constrained variational problem \eqref{constraint}, \eqref{dualfunctional} of Theorem \ref{thm:GLmain} corresponds to a classical obstacle problem (as is well-known), while in three (or higher) dimensions one may interpret it as a generalized, nonlocal, vectorial obstacle problem. 
Remark also that  norms related to $\|\cdot \|_*$ have been studied in the context of critical Sobolev spaces (see \cite{BBM,VS}).
\end{remark}

\begin{remark}Condition \eqref{Bbc} is easily seen to imply that $\supp(d^*B)\subset \bar\Omega$.
The converse holds if and only if $\Omega$ is simply connected. We do not know whether 
the minimizer of $\calE_0$ in the space  $\calC'' := \{ B\in H^1(\Lambda^2\R^3)\cap d\dot H^1(\Lambda^1\R^3) :\supp(d^* B)\subset \bar \Omega\}$ coincides with $B_*$ when
$\Omega$ fails to be simply connected.
\end{remark}

\begin{remark} It may appear unsettling that the functional $\calE_0$ contains the
non-gauge-invariant quantity $A_{ex}$. This can be effectively eliminated, however,
by decomposing $A_{ex}|_\Omega = A_{ex}^1+ A_{ex}^2$, where $A_{ex}^1\in (\ker d)^\perp$
 and $A_{ex}^2\in \ker d$ (see  \eqref{kerd.notation}, \eqref{kerd.char}). Then $A_{ex}^1$ is gauge-invariant, and
$\int_\Omega |d^*B + A_{ex}|^2 = 
\int_\Omega |d^*B + A^1_{ex}|^2 +
|A^2_{ex}|^2$.
\end{remark}

\bigskip
As mentioned before, a rather complete analysis of the asymptotic behavior of the Ginzburg-Landau functional for superconductivity in 2d can be found in \cite{SandSerfbook}. 
In the 3d case, some results in agreement with Proposition \ref{thm:3} have been obtained by formal arguments, as in \cite{Chapman}, and, rigorously in \cite{AlamaBronsardMontero} for the case of the ball, using some arguments of  \cite{JerrardMoteroSternberg}. In particular, 
\cite{AlamaBronsardMontero} identifies a candidate for the first critical field for the ball in a uniform applied field.%, and similar arguments could identify an analog of this candidate for more general domains. 
Our results show that this candidate first critical field is in fact correct. (In fact, it agrees exactly with the alternate expression for the critical field alluded to in Remark \ref{R1}, see Theorem \ref{thm:GLcritical}.)
Critical fields on thin superconducting shells, among other results, have been derived in \cite{Contreras,SternbergContreras} via a reduction to a limiting problem on a 2d manifold.

\subsection{the Gross-Pitaevsky functional}\label{sect:GPintro}

The second main object of study in this paper is a 
variational problem that describes  a Bose-Einstein
condensate with mass $m$, confined by a smooth potential $a:\R^3\to [0,\infty)$ such
that 
\beq\label{a.coercive}
a\in C^\infty(\R^3),
\quad\quad\quad\quad
\mbox{$a(x)\to +\infty$ as $|x|\to +\infty$,}
\eeq
and subjected to forcing $\Phi_\ep$ that in general depends on a scaling parameter $\e$. 
In the model case corresponding to rotation about the $z$-axis, 
$\Phi_\e :=  \frac 12 c_\e (x_1dx^2  - x_2dx^1)$, and $a(x)$ grows quadratically or faster. 
In this situation,
a stable condensate may described by  wave function $u_0$
that is a minimizer, local or global, in a function space
to be specified shortly, of a functional that may be written in the form
\beq\label{Gep.def}
\calG_\e(u) 
\ = \
\int_{\R^3} \frac 12 |\nabla u|^2 - \Phi_\e\cdot ju + \frac 1{4\e^2}(\rho -|u|^2)^2 + \frac w {2\e^2}|u|^2,
\eeq
where $j(u) = \frac i2 (  u d \bar u - \bar u du)$.
The functions $\rho,w$ appearing in the functional $\calG_\ep$ are determined by
the trapping potential $a$ and the mass $m$ as follows:
\beq\label{rho.def}
 \rho(x): = (\lambda - a(x))^+,
\quad w(x) :=  (\lambda - a(x))^-,
 \quad\ \mbox{for $\lambda$ such that }\int_{\R^3} \rho \ dx = m.
\eeq
The last condition clearly determines   $\lambda$ uniquely.

We study $\calG_\e$ in the function space 
\beq
H^1_a(\R^3;\C) := H^1_a := \mbox{ completion of }C^\infty_c(\R^3;\C)\mbox{ with respect to 
 }\| \cdot \|_a,
\label{H1a.def}\eeq
where the norm $\| \cdot \|_a$ is defined by $\| u\|_a^2 := \int_{\R^3} |du|^2 + (1+ a)|u|^2$. We also define
\[
H^1_{a,m}(\R^3;\C) := H^1_{a,m}  := \{ u\in H^1_a\ : \  \int |u|^2 = m\}.
\]
We will study the behavior of minimizers of $\calG_\e$ in $H^1_{a,m}$.

Throughout our discussion of the Gross-Pitaevsky functional we will  use the notation
\beq\label{Omega.def}
\Omega = \{ x\in \R^3 :\rho(x)>0\}.
\eeq
We will always assume that $\lambda$ is a regular value of 
$a$, so that $|Da|\ge c>0$ on $\partial \Omega$,  and hence
$w>0$ in $\R^3\setminus \bar \Omega$, and 
\beq\label{lambda.reg}
|\nabla\rho(x)|^2 + \rho(x) \ge c >0,
\quad
\ \ \rho(x) \ge c \dist(x,\partial \Omega), \ \ \quad \mbox{ for all }x\in \Omega.
\eeq

\subsubsection{results: Bose-Einstein condensates}

Our results for the Gross-Pitaevsky functional parallel those we obtain for the
Ginzburg-Landau functional: we identify a limiting variational problem, see \eqref{G0.def},
\eqref{GP.limspace} below, 
characterize when minimizers of the limiting problem are vortex-free,
and obtain a description of minimizers of the limiting problem as
solutions of a sort of nonlocal vector-valued obstacle problem.

We start by proving a theorem that characterizes $\Gamma$-limits of
the Gross-Pitaevsky functional, see Theorem \ref{GP.Gammalim}
in Section \ref{sect:GP} . This is parallel to Theorem 4 from \cite{BJOS1}
for the Ginzburg-Landau functional, and the proof relies on
results from \cite{BJOS1}
on the reduced GL functional. An immediate consequence
of Theorem  \ref{GP.Gammalim} is the following.

\begin{proposition}\label{prop:GPcompact}
Assume that 
$\Phi_\e = \logeps \Phi$, with $\Phi\in L^4_{loc}(\Lambda^1\R^3)$
and that $|\Phi(x)|^2 \le C a(x)$ outside some compact set $K$.

Assume that $u_\e$ minimizes $\calG_\e$ in $H^1_{a,m}$.
Then
\[
|u_\e| \to \rho\quad\mbox{ in }L^4(\R^3)
\]
for $\rho$ defined in \eqref{rho.def}, and there exists $j_0\in L^{4/3}(\Lambda^1\Omega)$ such that
\[
\logeps^{-1}ju_\e \rightharpoonup j_0\mbox{ weakly in }L^{4/3}(\R^3) %\mbox{ for  certain }p>1
\]
Moreover, $ j_0 = \rho v_0$,
where $v_0$ is the unique minimizer of 
\beq
\calG(v) := \int_{\Omega} \rho \left( \frac{|v|^2}2  - v \cdot \Phi  +\frac{1}{2} |d v|\right).
\label{G0.def}\eeq
in the space
\beq\label{GP.limspace}  
L^2_\rho(\Lambda^1\Omega) :=
\left \{ v \in L^1_{loc}(\Lambda^1\Omega)  \ : \ \int_{\Omega} \rho |v|^2 \ dx <\infty \right \}.
\eeq
(We set $\calG(v)=+\infty$ if $dv$ is not a Radon measure or if $\rho$ is not $|dv|$-integrable.)
\end{proposition}

\begin{remark}We also establish parallel results in the case when  $\Phi_\ep = \sqrt{g_\ep} \Phi$ with $\logeps \ll g_\ep \ll \ep^{-2}$. In particular, in this case the limiting energy
corresponding to \eqref{G0.def} is given by $\tilde G(v) = \int_{\Omega} \rho (|v-\Phi|^2 - |\Phi|^2)$.
See Remark \ref{rem:higher} for details.

It is clear that $v_0 =\Phi|_\Omega$ is the unique minimizer of $\tilde \calG$.
In particular, this implies that for a supercritical rotation around the (vertical) $x_3$ axis, 
corresponding to
$\Phi_\ep=\sqrt{g_\ep}\frac{c_0}{2}(x_1dx_2-x_2dx_1)$
with $g_\ep$ as above,
the limiting  ground-state vorticity is given by $dv_0 = c_0 dx_1 \wedge dx_2$, 
corresponding to an asymptotically uniform distribution of vortex lines throughout the condensate,
regardless of its geometry or topology.
This generalizes to 3 dimensions results obtained in \cite{CorrYng2008} in the 2d case.
\label{rem:higher1}\end{remark}

%Notice that in view of Remark \ref{rem:thm4}, we may assume $\Phi\in L^2_\rho(\Lambda^1\Omega)$ (instead of $\Phi$ smooth) in the statement of Proposition \ref{prop:GPcompact}.
%{\bf adjust regularity conditions on $a, \Phi$? perhaps we should either simply take both to be smooth, or else figure out for both what is really needed. My assumptions on $\Phi$
%are taken from Montero's paper, but I think we need less here.}

We next identify a necessary and sufficient condition
on $\Phi$ and $\rho$
for minimizers of the limiting functional $\calG$ to be vortex-free, by which we mean
that $dv_0 = 0$ in $\Omega$. 
For this result, it is useful to note that the space $L^2_\rho(\Lambda^1\Omega)$ defined
above is a Hilbert space with an inner product  and norm that we will write as
\[
(v,w)_\rho := \int_{\Omega} \rho\, v\cdot w \, dx, 
\quad\quad
\| v\|_\rho := (v,v)_\rho^{1/2}.
\]
We will sometimes use the same notation to denote the $\rho$-weighted $L^2$ inner product  or norm
for $k$-forms with values of $k$ other than $1$; the meaning should always be clear from the context.
We let $P_\rho$ denote the orthogonal projection with respect to the $L^2_\rho$ inner product, 
onto $(\ker d)_\rho$, where
\[
(\ker d)_\rho := L^2_\rho\mbox{-closure of } \{ \phi\in C^\infty(\Lambda^1\Omega) : d\phi = 0, \ \|\phi\|_\rho<\infty\}.
\]
We will also write $P_\rho^\perp$ for the complementary orthogonal projection.
Note that if $w\in \operatorname{Image}(P_\rho^\perp) = (\ker d)_\rho^\perp$, then $\int (\rho w)\cdot \phi = 0$
for all $\phi\in (\ker d)_\rho  \supset \ker d$.
Thus $\rho w \in (\ker d)^\perp$, and so it follows from the standard unweighted Hodge decomposition (see section  \ref{sect:hodge}, and in particular \eqref{hodgeperp}) that
\beq\label{kerdperprho}
\forall \ w\in (\ker d)_\rho^\perp, \ \exists \beta\in H^1_N(\Lambda^2\Omega)\mbox{ such that }w = \frac {d^*\beta}\rho
\mbox{ and  }
 \int _\Omega \frac{|d^*\beta|^2}\rho \ = \ \| w\|_\rho^2. %  < \infty.
\eeq
Thus if $\Phi\in L^2_\rho$, there exists $\beta_\Phi\in H^1_N$ such that 
$d^*\beta_\Phi\in L^2_\rho$ and 
$\Phi = P_\rho \Phi + \frac {d^*\beta_\Phi}\rho$.

We now state

\begin{theorem}
Suppose that $\Omega$ is a bounded, open subset of $\R^3$
and that $\rho\in C^1(\Omega)$ and $\Phi\in L^4_{loc}(\Lambda^1\R^3)\cap L^2_\rho(\Lambda^1\Omega)$  
are given, with $\rho$ satisfying \eqref{lambda.reg}.

Let $\beta_\Phi\in H^1_N(\Lambda^2\Omega)$ be such that $P^\perp_\rho \Phi =
\frac {d^*\beta_\Phi}\rho$, and let
 $\beta_0$ minimize the functional
\beq
\beta\mapsto  \frac 12 \int_{\Omega} \frac{|d^*\beta|^2}\rho
\label{GP.conmin1}\eeq
in the space
\beq
\left\{ \beta \in H^1_N(\Lambda^2\Omega) \ : \   \frac{d^*\beta}\rho \in L^2_\rho(\Lambda^1\Omega), \  \ \| \beta  - \beta_\Phi  \|_{\rho*}  \le \frac 12 \right\},
\label{GP.conmin2}\eeq
where
\beq
\| \beta\|_{\rho *} := 
\sup \left\{  \int_{\Omega} \beta \cdot dw : \ w\in C^\infty(\Lambda^1\bar\Omega),  \int_{\Omega} \rho |d w|\le 1\right\}.
\label{GP.conmin3}\eeq

Then $v_0 = P_\rho \Phi + \frac{d^*\beta_0}\rho$ is the unique minimizer of 
$\calG(\cdot)$ in $L^2_{\rho}(\Lambda^1\Omega)$.

Moreover,
\beq\label{eq:saturate}
\int_\Omega (\beta_\Phi-\beta_0)\cdot dv_0=\frac{1}{2}\int_\Omega\rho|dv_0|\, .
\eeq
Finally, 
$dv_0 = 0$ if and only if
$\| \beta_\Phi\|_{\rho *} \le \frac 12$.
\label{thm:GPcritical}\end{theorem}
Note that \eqref{eq:saturate} states that the action of the vorticity distribution $dv_0$   on the potential $\beta_0-\beta_\Phi$ is the largest possible given the constraint \eqref{GP.conmin2}. Similar considerations apply to \eqref{eq:B0dv0} and \eqref{constraint} in the case of superconductivity.

%Also add something about
%Higher rotations: equidistribution of vortices if $\Phi_\ep =  \frac 12 c_\ep (x_1dx^2 - x_2dx^1)$ with
%$\logeps \ll c_\ep \ll \e^{-1}$. 

We finally remark that in general
there can exist $\beta\in H^1_N(\Lambda^2\Omega)$ satisfying
$\frac{d^*\beta}{\rho}\in L^2_\rho(\Lambda^1\Omega)$ and  $\| \beta\|_{\rho*} <\infty$, but 
such that $\beta$ is not $|dw|$-integrable for some $w\in L^2_\rho(\Lambda^1\Omega)$
such that $\int \rho|dw|< 1$. Hence the restriction to smooth $1$-forms $w$
in the supremum that appears in the definition of the $\| \cdot \|_{\rho*}$ norm.

\subsubsection{related results about 2d BEC}

We do not know of any source in the literature that establishes 2d results analogous to  Proposition \ref{prop:GPcompact} and
Theorem \ref{thm:GPcritical}.  Such results are however in some sense known, at least as folklore, and can be established by arguing {\em exactly} as in the proofs
we supply here in the 3d case, but taking as a starting-point results from \cite{JS2002b}
about $\Gamma$-limits of the reduced Ginzburg-Landau functional in 2d, rather than the
analogous results about the same problem in 3d from \cite{BJOS1}, which (together with very general convex duality arguments) are the chief input in the relevant proofs. 

In particular, limits of
sequences of minimizers in 2d are described, in the same sense as
in Proposition \ref{prop:GPcompact}, by a functional $\calG$ on $L^2_\rho(\Omega')$ 
of {\em exactly} the same form 
as in \eqref{G0.def}, for a suitable $\Omega'\subset \R^2$. More generally, this functional can be obtained as a $\Gamma$-limit of the scaled 2d Gross-Pitaevsky energy, 
completely parallel to Theorem \ref{GP.Gammalim}. Moreover, this limiting functional
admits a dual formulation as functional with constraints, parallel to that in Theorem \ref{thm:GPcritical}, and from this one can easily determine a necessary and sufficient condition for the limiting vorticity to vanish. 

On the other hand, for more extreme rotation regimes in anharmonic trapping potentials in 2d, a quite detailed analysis has been carried out recently in \cite{CorrYng2008, CYR, R}. 

In a different direction, the critical rotation has been derived in certain higly symmetric domains in for example \cite{IM1, IM2, AAB}. These references also examine the behavior of minimizers for {\em slightly} supercritical rotations.

The main difference between 2 and 3 dimensions is the form of the constraint in the limiting variational
problem. In particular, in 2d, as in 3d, it is the case that if $v_0$ minimizes $\calG$, then
$dv_0 = d(\frac {d^*\beta_0}\rho)$, where the potential  $\beta_0$ minimizes the
functional \eqref{GP.conmin1}, subject to the constraint \eqref{GP.conmin2}, where the 
norm in the constraint is defined as in \eqref{GP.conmin3}. The difference is that in 2d,
the potentials $\beta$ are $2$-forms on $\R^2$, and so can be identified with functions.
And since  it is not hard to check that 
$
\{ d\omega : \int_\Omega \rho |d\omega| \le 1\}
$
is weakly dense in the set of signed measures $\mu$ such that $\int_\Omega \rho d|\mu|\le 1\}$,
the 2d constrained problem reduces to minimizing \eqref{GP.conmin1} in the set
\beq
\left\{  \beta \in H^1(\Lambda^2\Omega) : \| \frac 1 \rho(\beta - \beta_\Phi)\|_{L^\infty} \le \frac 12 \right\}.
\label{GP.conmin3bis}\eeq
This is a classical (weighted) 2-sided obstacle problem; for many $\Phi$, using the maximum principle
it in fact reduces to a one-sided obstacle problem.

{Thus we view the problem in Theorem \ref{thm:GPcritical} as a nonlocal, vector-valued analog of the 
classical obstacle problem.}

\subsubsection{previous work in 3d}

For  trapping potentials of the form $a(x) = \sum_{i=1}^3 \omega_i x_i^2$
with $\omega_i>0$, and
for $\Phi_\ep = \logeps\Phi  = \lambda \logeps( x_1dx_2-x_2dx_1)$,
the description of the critical rotation given in Theorem
\ref{thm:GPcritical}
was obtained in \cite{JerrardBEC}, building on earlier work
of \cite{AftRiv}. Indeed, these results show that if
$\| \beta_{\Phi}\|_* <\frac 12$, then
vorticity vanishes in the sense that
$Jv_\e\to 0$ as $\e\to 0$.
This is stronger than the estimate $\logeps^{-1}Jv_\e\to 0$ that follows
from Proposition \ref{prop:GPcompact} and  Theorem \ref{thm:GPcritical}.
The same paper \cite{JerrardBEC} characterized the $\Gamma$-limit of the
Gross-Pitaevsky functionals
for the particular choice of $a$ and $\Phi_\ep$ described above, in cases
where
one has an a priori bound on the part of the energy associated with the
vorticity of the condensate.
A similar $\Gamma$--limit result was shown  by Montero \cite{Montero2008}
to hold
for very general trapping potentials $a$ and forcing terms $\Phi_\ep$.
Although it was not done in \cite{Montero2008}, this result could in
principle
be used to prove that $Jv_\e\to 0$ for subcritical rotations in these situations.
The a priori bound
on part of the energy required for these results means that they cannot
give any information
about minimizers for supercritical rotations.

\bigskip
\noindent{\bf Acknowledgments.} S.B., R.L.J. and G.O. wish to thank FIM at ETH Z\"urich, where part of this work was completed, for the warm hospitality.

\section{background and notation}

\subsection{differential forms}\label{sect:diff_forms}

If $U$ is an open subset of $\R^n$, we will use the notation $W^{1,p}(\Lambda^k U)$
to denote the space of maps $U\to \Lambda^k\R^n$ (that is, $k$-forms on $U$) 
that belong to the Sobolev space $W^{1,p}$. A generic element $\omega\in W^{1,p}(\Lambda^k U)$ thus has the form
\beq\label{generic.form}
\sum_{\{ \alpha : 1\le \alpha_1<\ldots<\alpha_k\le n\} } \omega_\alpha dx^{\alpha_1}\wedge \cdots \wedge dx^{\alpha_k}
\eeq
with $\omega_\alpha\in W^{1,p}(\Omega;\R)$ for every multiindex $\alpha$. 
We use the notation $L^p(\Lambda^k U)$,
$C^\infty(\Lambda^kU)$, and so on in a parallel way.

For an open set $\Omega$ with nonempty boundary
and $\omega\in C^0(\Lambda^k \bar \Omega)$, we define $\omega_\top$ and
$\omega_N$
in $C^0(\Lambda^k\partial \Omega)$
by 
\[
\omega_\top = i^*\omega, \mbox{ where $i:\partial \Omega \to \bar \Omega$ is the inclusion};
\quad
\quad
\quad
\omega_N = \omega|_{\partial \Omega} - \omega_\top.
\]
One refers to $\omega_\top$ and $\omega_N$ as the tangential and normal
parts of $\omega$ on $\partial \Omega$. We will use the same notation
$\omega_\top, \omega_N$ to refer to the tangential and normal
parts of (the trace of) a form $\omega\in W^{1,p}(\Lambda^k\Omega)$, which
one can define by noting that for example the
map $\omega\mapsto \omega_\top$, well-defined on a dense subset of $W^{1,p}(\Lambda^k\Omega)$,
extends to a bounded linear map $W^{1,p}(\Lambda^k\Omega)\to L^{p}(\Lambda^k\partial\Omega)$,
or equivalently by applying the pointwise definition of $\omega_\top$, say, to the trace of $\omega$ at 
a.e. point of $\partial \Omega$.

If $\omega, \phi$ are elements of $L^2(\Omega;\Lambda^k\R^n)$, written as in \eqref{generic.form}, we will write $\omega\cdot\phi$
to denote the integrable function defined by
\[
\omega \cdot \phi = \sum_{\{ \alpha : 1\le \alpha_1<\ldots<\alpha_k\le n\} } \omega_\alpha \phi_\alpha.
\]
This allows us to define an $L^2$ inner product on spaces of differential forms in the obvious way.
We write $d^*$ to denote the formal adjoint of $d$, so that $\int d\omega\cdot \phi = \int\omega\cdot d^*\phi$
when $\omega$ is a smooth $k-1$-form and $\phi$ a smooth $k$-form for some $k$, and at least one of them has compact support. Then
\[
d^* \phi = (-1)^{k} \star d\star \phi \quad\mbox{ if $\phi$ is a $k$-form},
\]
where in $\R^3$, the $\star$ operator, mapping $k$-forms to $(3-k)$-forms,  is characterized by
\[
\omega \wedge \star \phi = \star \omega \wedge \phi = \omega\cdot \phi  \ dx^1\wedge dx^2\wedge dx^3.
\]
(In even dimensions one must be more careful about signs.)

We will use the notation
\begin{align*}
W^{1,p}_\top  (\Lambda^k\Omega)
&:= \{ \omega\in W^{1,p}(\Lambda^k\Omega) : \omega_\top = 0 \},\\
W^{1,p}_N  (\Lambda^k\Omega)
&:= \{ \omega\in W^{1,p}(\Lambda^k\Omega) : \omega_N = 0 \}.
\end{align*}
and
\begin{align*}
\calH_\top  (\Lambda^k\Omega)
&:= \{ \omega\in W^{1,p}(\Lambda^k\Omega) : \omega_\top = 0, d\omega  = d^*\omega = 0 \},\\
\calH_N  (\Lambda^k\Omega)
&:= \{ \omega\in W^{1,p}(\Lambda^k\Omega) : \omega_N = 0, d\omega  = d^*\omega = 0  \}.
\end{align*}
In fact forms in $\calH^k_{\top}$ and  $\calH^k_{N}$ are known to be smooth.

 Gauge-invariance implies that the set of minimizers of $\calF_\e$ is noncompact  in
 $H^1(\Omega;\C)\times [A_{ex}+H^1(\Lambda^1\R^3)]$. In order to remedy this, we will often restrict $\calF_\e$ to a smaller space.
Thus
we introduce 
\beq\label{H1star.def}
\dot H^1_{*} (\R^3; \Lambda^1\R^3) =  \dot H^1_* :=  \{ A\in \dot H^1(\R^3;\Lambda^1\R^3) : d^* A = 0 \}
\eeq
with the inner product  $(A,B)_{\dot H^1_{*} (\Lambda^1\R^3)} = (A,B)_{*}:= (dA, dB)_{L^2(\Lambda^2\R^3) }$.
This makes $\dot H^1_{*} (\Lambda^1\R^3)$ into a Hilbert space, 
satisfying in addition the Sobolev inequality
\[
\| A \|_{L^6(\Lambda^1\R^3)} \le C \| A \|_{\dot H^1_{*}(\Lambda^1\R^3)} \, .
\]
In view of standard results about the Hodge decomposition,
given any any 1-form $\tilde A$ such that $\tilde A \in  A_{\e,ex}+ \dot H^1(\R^3; \Lambda^1\R^3)$, we can write $\tilde B : =  \tilde A  - A_{\e,ex} \in \dot H^1$ in the form
\[
\tilde B = B + d\phi, \quad\quad\mbox{ where }B\in H^1_*\ \mbox{ and }d\phi\in \dot H^1(\R^3;\Lambda^1\R^3).
\]
Thus given any pair $(\tilde u, \tilde A)\in H^1(\Omega;\C)\times [A_{ex}+H^1(\Lambda^1\R^3)]$, there exists an equivalent pair $(u, A) = (\tilde u  e^{-i\phi}, \tilde A - d\phi)$  in $H^1(\Omega;\C)\times [ A_{\e,ex}+\dot H^1_*]$, so that in restricting $\calF_\e$ to
$H^1(\Omega;\C)\times [ A_{\e,ex}+\dot H^1_*]$, we do not sacrifice any generality.

\subsection{Hodge decompositions}\label{sect:hodge}

We will need several Hodge decompositions.
First, on a bounded open domain $\Omega$ with $C^1$ boundary, 
we have, for every integer $k\in \{0,\ldots, n\}$ the decompositions
\beq
L^2(\Lambda^k\Omega) = d H^1(\Lambda^{k-1}\Omega)\oplus d^* H^1_N(\Lambda^{k+1}\Omega) \oplus  \calH_N (\Lambda^k\Omega)
\label{Ndecomp}\eeq
and 
\beq
L^2(\Lambda^k\Omega) = d H^1_\top(\Lambda^{k-1}\Omega)\oplus d^* H^1(\Lambda^{k+1}\Omega) \oplus  \calH_\top (\Lambda^k\Omega).
\label{Tdecomp}\eeq
These are  known from work of Morrey (see also \cite{ISS}, Theorem 5.7).
The first of these, for example, means that every $\omega\in L^2(\Lambda^k\Omega)$ can be written in the form
$
\omega = d\alpha + d^*\beta + \gamma$,
where $\alpha\in H^1(\Lambda^{k-1}\Omega), \beta \in d^* H^1_N(\Lambda^{k+1}\Omega)$,
and $\gamma\in \calH^k_N$, and moreover  $d\alpha, d^*\beta$, and $\gamma$ are mutually
orthogonal in $L^2$.  

We will sometimes use the notation
\beq
\ker d = H^1(\Omega) \oplus \calH_N(\Lambda^1\Omega), \quad\quad
(\ker d)^\perp = d^*H^1_N(\Lambda^2\Omega).
\label{kerd.notation}\eeq
This is justified by the following considerations. First,
we claim that for $v\in L^2(\Lambda^1\Omega)$,
\beq
dv = 0 \mbox{ as a distribution on }\Omega \quad \iff 
v\in dH^1(\Omega) \oplus \calH_N(\Lambda^1\Omega).
\label{kerd.char}\eeq
Indeed,  to prove that $v\in dH^1(\Omega) \oplus \calH_N(\Lambda^1\Omega)$, it
suffices by \eqref{Ndecomp} to verify that $v \perp d^*H^1_N(\Lambda^2\Omega)$. 
Fix any $\beta\in H^1_N(\Lambda^2\Omega)$, and let $\chi_\ep\in C^\infty_c(\Omega)$ be a sequence
of functions such that $\chi_\ep = 1$ in $\{ x\in \Omega : \dist(x,\partial \Omega) > \ep\}$,
$\|\nabla\chi_\ep\|_\infty\le C\ep$. Then  the assumption that $dv=0$ in $\Omega$ implies that
\[
0 = \int_\Omega v \cdot d^*(\chi_\ep \beta) = 
\int_\Omega \chi_\ep  v \cdot d^* \beta 
+
\int_\Omega  v \cdot \star  \cdot ( d\chi_\ep \wedge \star \beta)
\]
for every $\ep$. Thus
\[
\int_\Omega   v \cdot d^* \beta = 
\lim_{\ep\to 0}
\int_\Omega \chi_\ep  v \cdot d^* \beta 
= -\lim_{\ep\to 0}
\int_\Omega v \cdot \star ( d\chi_\ep \wedge \star \beta) = 0
\]
where the last equality follows from the fact that $\beta_N = 0$.
This proves one implication in \eqref{kerd.char}, and the other is obvious.

Similarly, the Hodge decomposition implies that if $\omega\in  L^2(\Lambda^k\Omega)$, then 
\beq\label{hodgeperp}
\omega\in d^*H^1_N(\Lambda^{k+1}\Omega)
\quad\iff
\quad\int_\Omega \omega\cdot \phi  = 0  \ \ \  \forall \ \phi\in H^1(\Lambda^{k}\Omega)
\mbox{ with }d\phi = 0 ,%\} = d^*H^1_N(\Lambda^k\Omega).
\eeq
%\beq\label{hodgeperpbis}
%\omega\in dH^1_\top(\Lambda^{k-1}\Omega)
%\quad\iff
%\quad\int_\Omega \omega\cdot \phi  = 0  \ \ \ \forall \  \phi\in H^1(\Lambda^{k}\Omega)
%\mbox{ with }d^*\phi = 0 .%\} = d^*H^1_N(\Lambda^k\Omega).
%\eeq

%Thus we will sometimes write for example $(\ker d)^\perp:=  d^* W^{1,2}_N(\Lambda^{k+1}\Omega)$, where $k$ should be clear from the context. 

We also define
\beq
P := {L^2(\Lambda^1\Omega)} \mbox{ orthogonal projection onto }
d^* W^{1,2}_N(\Lambda^2 \Omega) = (\ker d)^\perp.
\label{morrey1}\eeq
Given $A\in \dot H^1(\Lambda^1\R^3)$ for example, we will generally abuse notation and
write $PA_1$ instead of $P( A_1|_\Omega)$.
We remark that 
\beq
\| PB\|_{L^2(\Omega)}^2 = \inf \{  \|B + \gamma \|_{L^2(\Omega)}^2 \ : \gamma\in H^1(\Omega), d\gamma=0  \}.
\label{morrey}\eeq
%We also note that for $\beta\in H^1(\Lambda^2\R^3)$,
%\beq
%\mbox{ if }\supp d^*\beta \subset \bar \Omega, \quad\mbox{ then }\beta_N = 0\mbox{ on }\partial \Omega.
%\label{implicitBC}\eeq
%Indeed, if $\supp d^*\beta\subset \bar \Omega$ and $f\in H^1(\R^3)$ has compact support, then
%\[
%\int_\Omega d^*\beta \cdot df = \int_{\R^3} d^*\beta \cdot df = 0.
%\]
%By standard extension theorems, any $f\in H^1(\Omega)$ can be written as the restriction to
%$\Omega$ of a compactly supported function in $H^1(\R^3)$, we conclude that 
%$d^*\beta \perp dH^1(\Omega)$, and hence, from one of the Hodge decompositions above, that
%$d^*\beta \in d^* H^1_N(\Lambda^2\R^3) + \mathcal{H}_N(\Lambda^1\R^3)$.
%In particular, 

For applications to Bose-Einstein condensates we will need a Hodge decomposition in the weighted Hilbert space $L^2_\rho(\Lambda^k\Omega)$. In particular, in the notation from the introduction (compare \eqref{GP.limspace}, \eqref{kerdperprho}), we may decompose\footnote{Notice that our notation is inconsistent, with $P_\rho$ = projection onto $(\ker d)_\rho$ for Bose-Einstein,  
and $P$ := projection onto $(\ker d)^\perp$ for superconductivity. These
conventions are convenient however, and we do not think they can lead to any confusion.}
$\phi\in L^2_\rho(\Lambda^1\Omega)$ as
$$
\phi=\omega+\frac{d^*\beta}{\rho},\qquad\omega = P_\rho\phi\in(\ker d)_\rho, \qquad
\frac{d^*\beta}{\rho}\in L^2_\rho(\Lambda^1\Omega).
$$
For refined results assuming $\rho$ and $\phi$ sufficiently smooth, see \cite{Montero2008}.

\subsection{duality}

We will frequently use the following basic result, whose proof can be found for example in
\cite{ET}.

\begin{lemma} Assume that $H$ is a Hilbert space, and that
$I:H\to (-\infty, \infty]$ is a convex function and that $I(x)<\infty$ for some $x\in H$.

Let $G(x) := I(x) + \frac 12 \| x\|_H^2$.

Let $I^*$ denote the Legendre-Fenchel transform of $I$, so that
\[
I^*(\xi) := \sup_{x\in H}\left( (\xi, x)_H - I(x)\right).
\]
Then if we define $G^\dagger(x) := I^*(-x) +\frac 12 \|x \|_H^2$, the following hold:

\begin{enumerate}
\item There exists a unique $x_0\in H$ such that $G(x_0) = \min_H G(\cdot)$.
\item The same $x_0\in H$ is the unique minimizer of $G^\dagger$ in $H$.
\item $G(x) + G^\dagger(y) \ge 0$, and $G(x) + G(y) = 0$ if and only if $(x,y) = (x_0, x_0)$.
\end{enumerate}
\label{lem:dual}\end{lemma}

\section{vortex density in 3d superconductors}

\subsection{A dual variational problem}

We start with the proof of Theorem \ref{thm:GLmain}, in which we identify a variational problem
dual to that of miminizing $\calF$, which describes
the limiting density of vortex lines in a superconducting material subjected to an applied magnetic field. We then use this dual problem to prove Theorem \ref{cor:critfield},  giving a
necessary and sufficient condition for the limiting vorticity to vanish. %Similar arguments have been given in 2 dimensions, and so in a sense the main mathematical novelty here is found in our compacnion paper \cite{BJOS1}, which justifies the study of the functional $\calF$. 

In the next section we present several different and, actually,  simpler derivations of (an equivalent but different-looking expression for) the critical field. The approach presented here, although a little more complicated, has the advantage of yielding the dual problem of the statement of Theorem \ref{thm:GLmain}, which clearly generalizes, 
in an interesting way, the obstacle problem identified in the 2d literature, 
see \cite{SandSerfbook}.

\begin{proof}[Proof of Theorem \ref{thm:GLmain}]
{\bf Step 0}. 
Let us write $\xi = A|_\Omega -v$ and $\zeta = A - A_{ex}$, so that in terms of the $\xi,\zeta$ variables,
\[
\calF(v,A) = \frac 12 \int_{\Omega} |\xi|^2 + |d(\zeta-\xi+A_{ex})| + \frac12\int_{\R^3}|d\zeta|^2 =: F(\xi, \zeta).
\]
Also, let $H := L^2(\Lambda^1\Omega)\times  \dot H^1_*$.
Note that $H$ is a Hilbert space with the norm
\[
\| (\xi ,\zeta )\|_H^2 := \| \xi \|_{L^2(\Omega)}^2 + \| d\zeta \|_{L^2(\R^3)}^2
\]
and the corresponding inner product. We next define
\[
I(\xi, \zeta ) :=  \frac 12 \int_\Omega |d(\zeta-\xi+A_{ex})| 
\]
so that 
\[
F(\xi, \zeta ) = \frac 12 \| ( \xi, \zeta  )\|_H^2 + I( \xi, \zeta ).
\]
As usual, $I$ is understood to equal $+\infty$ if $d( \zeta-\xi+A_{ex})$ is not a Radon measure.
Let $I^*$ denote the Legendre-Fenchel transform of $I$, so that
\[
I^*(\xi, \zeta) = \sup_{(\xi^*, \zeta)\in H} \big\{ ((\xi, \zeta), (\xi^*, \zeta^*))_H - I(\xi^*, \zeta^*) \big\}
\]
Let us further write 
\[
F^\dagger(\xi, \zeta) = \frac 12 \| (\xi, \zeta )\|_H^2 + I^*(\xi, \zeta).
\]

{\bf Step 1}.
As remarked in Lemma \ref{lem:dual} above, $(\xi_0, \zeta_0)$ minimizes $F$ 
if and only if $(-\xi_0, -\zeta_0)$ minimizes $F^\dagger$. 
To  compute $I^*$, note that for $(\xi, \zeta)\in H$, 
\begin{align}
I^*(\xi, \zeta) 
&\ =\   
\sup_{(\xi^*, \zeta^*)\in H} \left\{ ( (\xi, \zeta), ((\xi^*- A_{ex}) + A_{ex}, \zeta^*))_H -  \frac 12 \int_\Omega |d(\zeta^*-(\xi^*-A_{ex}))| \right\}
\nonumber\\
&\ =\   
(\xi, A_{ex})_{L^2(\Omega)} + \sup_{(\xi^*, \zeta^*)\in H} \left\{ ( (\xi, \zeta), (\xi^*, \zeta^*))_H - \frac 12 \int_\Omega |d (\zeta^*- \xi^*)| \right\}.
\label{d1}
\end{align}
It is clear  the supremum on the right-hand side equals zero 
if $(\xi, \zeta)$ satisfies
\beq
\int_{\R^3} \chara_\Omega \xi\cdot \xi^* + d\zeta\cdot d\zeta^* \  \le \frac 12 \int _\Omega |d (\zeta^*- \xi^*)| 
\quad\quad\mbox{ for all }(\xi^*, \zeta^*)\in H.
\label{d2}\eeq
and if this condition fails to hold, then  (by homogeneity) the sup in \eqref{d1} is infinite.
Thus
\[
I^*(\xi, \zeta) = \left\{
\begin{array}{ll}
(\xi,A_{ex})_{L^2(\Omega)}	\ \ &\mbox{ if \eqref{d2} holds}\\
+\infty 	&\mbox{ if not}.
\end{array}\right.
\]
It follows that
\[
F^\dagger(\xi, \zeta) = \left\{
\begin{array}{ll} 
\frac 12
%\left[\| w-A_{ex}\|_{L^2(\Omega)}^2 - \|A_{ex}\|_{L^2(\Omega)}^2 + \| db\|_{L^2(\R^3)}^2]\right]
\| (\xi +A_{ex}, \zeta)\|_{H}^2 - \frac 12 \|A_{ex}\|_{L^2(\Omega)}^2
\ \ \ &\mbox{ if \eqref{d2} holds}\\
+\infty 	&\mbox{ if not}.
\end{array}\right.
\]

{\bf Step 2}.
We want to rewrite $F^\dagger$ in a more useful form. To this end, we first claim that $(\xi,\zeta)\in H$
satisfies \eqref{d2} if and only if  %$\xi\in d^*H^1_N(\Lambda^1\Omega)$, and the following hold:
\begin{equation}
\int_{\R^3} d\zeta\cdot d\zeta^*  \ \ \le \ \ 
\frac 12 \int _\Omega| d \zeta^*| 
\quad \mbox{ for all } \zeta^*\in \dot H^1(\Lambda^1\R^3)
\label{d5a}\end{equation}
and 
\begin{equation}
\zeta\in H^2_{loc}\cap H^1_*, \quad\mbox{ and } d^*d\zeta + \chara_\Omega \xi = 0.
\label{d6aa}\end{equation}

{\bf Step 2a}.
First assume that \eqref{d2} holds. Note that
since $(\xi, \zeta)\in H$
\[
( (\xi, \zeta), (\xi^*,\zeta^*))_H \le \frac 12 \int _\Omega |d (\zeta^*- \xi^*)| 
\quad\quad\mbox{ for all }(\xi, \zeta)\in L^2(\Lambda^1\Omega)\times \dot H^1(\Lambda^1\R^3).
\]
This follows from \eqref{d2}, since we can write
$(\xi^*, \zeta^*)\in L^2(\Omega)\times \dot H^1(\R^3)$  as
$(\xi^*, \zeta')+(0,\zeta'')$ with $(\xi^*, \zeta')\in H$ and $\zeta'' \perp \dot H^1_*$, so that
$d\zeta'' \equiv 0$.

Now  we immediately obtain  \eqref{d5a} by taking
$(\xi^*, \zeta^*)$ of the form
$(0, \zeta^*)$ in the above inequality. Similarly, by 
choosing $(\xi^*, \zeta^*)$ of the form $\pm(\zeta^*|_\Omega, \zeta^*)$ we find that
\begin{equation}
\int_{\R^3} (d \zeta\cdot d\zeta^*  + \chara_\Omega \xi\cdot \zeta^*) = 0
\quad\quad\mbox{ for all }\zeta^*\in \dot H^1(\Lambda^1\R^3).
%\int_\Omega \xi \cdot \zeta  + \int_{\R^3} d\zeta \cdot d \zeta= 0 \quad \mbox{ for all } \zeta\in \dot H^1_*.
\label{d6a}\end{equation}
Since $d^* \zeta = 0$ for  all $\zeta \in \dot H^1_*$, we see from \eqref{d6a}
that $-\Delta \zeta + \chara_\Omega \xi = 0$ as distributions, and hence from elliptic regularity that 
$\zeta\in H^2_{loc}(\R^3)$ and that $d^*d\zeta+\chara_\Omega\xi = 0$
a.e. in $\R^3$, so that \eqref{d6aa} holds.

{\bf Step 2b}.
Conversely, suppose that \eqref{d5a}, \eqref{d6aa} hold.  Clearly \eqref{d6aa} implies \eqref{d6a}, so for $(X^*, \zeta^*)
\in \dot H^1(\Lambda^1\R^3)\times \dot H^1(\Lambda^1\R^3)$,
\begin{align*}
\int_\Omega\xi\cdot X^* \ + \int_{\R^3}d\zeta\cdot d\zeta^*
&\overset{ \eqref{d6a}}=
\int_{\R^3}  d\zeta\cdot d(\zeta^*-X^*)\\
&\overset{ \eqref{d5a}}\le
\frac 12\int_\Omega |d(\zeta^*-X^*)|.
\end{align*}
Thus \eqref{d2} follows whenever $\xi^*$ is the restriction
to $\Omega$ of some $X^* \in  \dot H^1(\Lambda^1\R^3)$.
 
We next deduce from this that \eqref{d2} holds whenever $\xi^*\in L^2(\Lambda^1\Omega)$.
We may assume that $d\xi^*$ is a measure, as otherwise the right-hand side of \eqref{d2} is infinite and there
is nothing to prove. Then, given $(\xi^*, \zeta^*)$, it suffices to find $(X_\e^*, \zeta_\e^*)\in \dot H^1(\Lambda^1\R^3)\times \dot H^1(\Lambda^1(\R^3)$ such that
\beq
\begin{aligned}
&X_\e^*|_\Omega \rightharpoonup \xi^* 
\mbox{  weakly in }L^2(\Lambda^1\Omega),\\
&d\zeta_\e^* \rightharpoonup d\zeta^*
\mbox{  weakly in }L^2(\Lambda^1\R^3),
\mbox{ and }\\&
\int_{\Omega}|d(\zeta_\e^*-X_\e^*)|\rightharpoonup \int_\Omega|d(\zeta^* -\xi^*)|.
\end{aligned}
\label{approx1}\eeq

To do this, we start by fixing, for all $\ep$ sufficiently small, a $C^1$ diffeomorphism
$\Psi_\ep:\R^3\to\R^3$ such that
\beq
\Psi_\ep(\{ x\in \R^3: \dist(x, \Omega)<\ep\})\subset \Omega, \quad\quad
\mbox{$\Psi^\ep(x)=x$
if $\dist(x, \partial\Omega) > \sqrt \e$ }
\label{Psiep1}\eeq
and
\beq
\| D\Psi_\ep - I\|_\infty \le C \sqrt\ep, \quad\quad\quad
\| D\Psi_\ep^{-1} - I\|_\infty \le C \sqrt\ep.
\label{Psiep2}\eeq
For example we may take $\Psi_\ep$ in $\{x\in \R^3: \dist(x, \partial\Omega) \le \sqrt \e\}$
to have the form $\Psi_\ep(s) = x - f_\ep(d(x)) \bar \nu(x)$,
where $\bar \nu(x)$ is the outer unit normal to $\partial \Omega$ at the point of $\partial \Omega$
closest to $x$, and $d(x)$ is the signed distance (positive outside $\Omega$, negative in $\Omega$) to $\partial \Omega$, and $f_\ep$ is a  nonnegative function with compact support in $(-\sqrt\ep, \sqrt \ep)$
such that $|f_\ep'| \le C \sqrt \ep$ and $f_\ep(\ep) > \ep$.

Next, let $\bar \xi^*$ denote some extension of $\xi$ to an element of $L^2(\Lambda^1\R^3)$,
and let $\psi_\ep$ be a smooth nonnegative radially symmetric mollifier with support in $B(0, \ep/2)$ and
such that $\int \phi_\ep -1$.

Then we define
 \[
 X_\e^* := \psi_\e *( \Psi_\e^\# \bar \xi^*), \quad\quad\mbox{ and } \ \ 
\zeta_\ep^* := \psi_\ep * (\Psi_\e^\# \zeta^*).
\]
Then the verification of  \eqref{approx1} follows by a reasonably straightforward, classical argument.
(See for example the proof of Lemma \ref{lem:GPdense}, at the end of Section \ref{S:becGlim}, where similar
computations are carried out in detail in a somewhat more complicated setting.)

{\bf Step 3}. 
In view of \eqref{d6aa}, 
we can eliminate $\xi$ from the expression for $F^\dagger$ to find that
\[
F^\dagger(\xi, \zeta) = \left\{
\begin{array}{ll} 
\frac 12
%\left[\| w-A_{ex}\|_{L^2(\Omega)}^2 - \|A_{ex}\|_{L^2(\Omega)}^2 + \| db\|_{L^2(\R^3)}^2]\right]
\|( - d^*d\zeta +A_{ex}, \zeta)\|_{H}^2 - \frac 12 \|A_{ex}\|_{L^2(\Omega)}^2
\ \ \ &\mbox{ if \eqref{d5a}, \eqref{d6aa} hold}\\
+\infty 	&\mbox{ if not}.
\end{array}\right.
\]

We now rewrite everything in terms of $A = \zeta+ A_{ex}$,\ 
$v = A|_\Omega - \xi =  (\zeta+ A_{ex})|_\Omega-\xi $,  and $B = d(A-A_{ex}) = d\zeta$.

First, the constraints \eqref{d5a}, \eqref{d6aa} are equivalent to
the conditions appearing under part {\bf 1} of the statement of the theorem,
that is, 
\beq\label{rewrite.constraints}
\mbox{$B \in \calC$, \ and  \ \ \  
$d^*B + \chara_\Omega(A-v) = 0$,}
\eeq
where the constraint set $\calC$ is defined in \eqref{constraint}.

Second, it follows from Lemma \ref{lem:dual} that  
\begin{align*}
&\mbox{$(v_0, A_0)$ minimizes $\calF$ in $L^2(\Omega)\times [A_{ex}+\calH_0]$}
\\
&\quad\quad\quad\quad
\iff
\mbox{$(A_0-v_0, A_0-A_{ex})$ minimizes $F$ in $H$}
\\
&\quad\quad\quad\quad
\iff
\mbox{$(v_0-A_0, A_{ex}  - A_0)$ minimizes $F^\dagger$ in $H$},
\\
&\quad\quad\quad\quad
\iff
\mbox{$(v_0,A_0)$ minimizes $\calF^\dagger$ in $L^2(\Omega)\times [A_{ex}+\calH_0]$},
\end{align*}
where $\calF^\dagger(v,A) := F^\dagger(v-A, A_{ex}-A) + \frac12\|A_{ex}\|_{L^2(\Omega)}^2 $.

Thus 
\begin{equation}
\calF^\dagger(v,A) = \left\{
\begin{array}{ll} 
\frac 12
%\left[\| w-A_{ex}\|_{L^2(\Omega)}^2 - \|A_{ex}\|_{L^2(\Omega)}^2 + \| db\|_{L^2(\R^3)}^2]\right]
\| (v-A+A_{ex}, A_{ex}-A)\|_{H}^2 
\ \ \ &\mbox{ if \eqref{rewrite.constraints} holds, and }\\
+\infty 	&\mbox{ if not}.
\end{array}\right.
\label{d3}
\end{equation} 
Rewriting $\calF^\dagger$ in terms of $B := d(A-A_{ex})$,
it follows that  $(v_0,A_0)$ minimizes  $\calF^\dagger$ if and only if
conditions {\bf 1} and {\bf 2} from the statement of the theorem hold.
Remark finally that \eqref{eq:B0dv0} follows from \eqref{v0eqn} and the relation
\beq\label{eq:varv0}
\frac{1}{2}\int_\Omega|dv_0| \ + \ 
\int_\Omega (v_0-A_0)\cdot v_0 = 0
\eeq
which in turn follows by stationarity of $\calF(v_0,A_0)$ with respect to variations $v_t=e^t\cdot v_0$ around $t=0$.
\end{proof}

\subsection{first characterization of the critical applied magnetic field}
We next want to prove Theorem \ref{cor:critfield}, which gives a 
necessary and sufficient condition
for the vorticity of a minimizer of $\calF$ to be nonzero.
Recall that this criterion involves the minimizer
of an energy $\calE_0$ in a space $\calC'$, defined in \eqref{constraint'}. 
We first establish
some facts about $\calC'$. Given a function $v$ defined on $\Omega$, we use the notation $\chara_\Omega v$ to denote its extension to the function, defined on $\R^3$, that vanishes away from $\Omega$.

\begin{lemma}
Assume that $B\in H^1(\Lambda^2\R^3)\cap d \dot H^1(\Lambda^1\R^3)$.

If $B\in \calC'$, then $\supp(d^*B)\subset \bar \Omega$, and $(d^*B)|_\Omega\in d^*H^1_N(\Lambda^2\Omega) = (\ker d)^\perp$.

Conversely, given any $\phi\in d^*H^1_N(\Lambda^2\Omega)$, there exists 
$B_\phi\in \calC'$ such that $d^*B_\phi = \chara_\Omega \phi$

Finally, 
\beq
\calC'\subset \bigcap_{1<p\le 2} \dot W^{1,p}  \ \  \subset \ \  
\bigcap_{\frac 32 < q \le 6} \dot L^{q}.
%  \ \ \subset \ \  \bigcap_{r>6}L^r,
\label{Creg}\eeq
\label{LCprime}\end{lemma}

\begin{remark}The proof will show that
$B_\phi = d(-\Delta)^{-1}(\chara_\Omega \phi)$, where $(-\Delta)^{-1}$ denotes
convolution with the fundamental solution for the Laplacian on $\R^3$, 
with $(-\Delta)^{-1}(\chara_\Omega\phi) \in \cap_{1<p\le 2} \dot W^{2,p} \subset \cap_{r>3} L^r$.
\label{Rpsi}\end{remark}

\begin{proof}
{\bf Step 1}. We first claim that if $B\in \calC'$, then
%First assume that $B\in \calC'$, which we recall means that
%$B\in H^1(\Lambda^2\R^3)\cap d\dot H^1(\Lambda^1\R^3) $ and that
\beq
\int_{\R^3} d^*B\cdot v = 0\quad\quad\mbox{ for all $v\in L^2(\Lambda^1\R^3)$ such that 
$dv=0$ in $\Omega$}.
\label{Cprime1}\eeq
(Recall that by definition of $\calC'$, this identity holds for $v\in H^1(\Lambda^1\R^3)$ such that 
$dv=0$ in $\Omega$.)
To see this, define  a diffeomorphism $\Psi_\ep:\R^3\to \R^3$ 
as in \eqref{Psiep1}, \eqref{Psiep2}, and let 
$\psi_\ep$ denote a symmetric approximate identity supported in $B(0,\ep/2)$.
Given $v\in L^2(\Lambda^1\R^3)$ such that $dv=0$ in $\Omega$,
define $v_\ep := \psi_\ep*(\Psi_\ep^\# v)\in H^1(\Lambda^1\R^3)$. 
Clearly $v_\ep\to v$ in $L^2(\R^3)$, as $\ep\to 0$, and we also claim that $dv_\ep = 0$
in $\Omega$. 
To see this, note that
for any $\phi\in C^1_c(\Lambda^2\Omega)$, 
\[
\int_\Omega dv_\ep \cdot \phi = \int_\Omega v_\ep \cdot d^*\phi
= 
\int_{\R^3} \Psi_\ep^\# v \cdot (\psi_\ep*d^*\phi)
=
\int_{\R^3} \Psi_\ep^\# v \wedge d \star(\psi_\ep*\phi).
%=
%\int_{\Psi_\ep(\Omega)}  v \cdot (\Psi_\ep^{-1})^\#\psi_\ep*d^*\phi
\] 
Since $ \Psi_\ep^\# v \wedge d \star(\psi_\ep*\phi) = \Psi_\ep^\#
[v \wedge ( \Psi_\ep^{-1})^\#(d \star(\psi_\ep*\phi)) ]$,
it follows that
\beq
\int_\Omega dv_\ep \cdot \phi 
= \int_{\Psi_\ep(\R^3)}v \wedge d  ( \Psi_\ep^{-1})^\#( \star(\psi_\ep*\phi)).
= \int_{\R^3}v \cdot d^* \star \phi_\ep
\label{mol.rewrite}\eeq
for $\phi_\ep := ( \Psi_\ep^{-1})^\#( \star(\psi_\ep*\phi))$.
The definitions of $\Psi_\ep$ and $\psi_\ep$ imply that $\phi_\ep$ has compact support in
$\Omega$. Thus $ \int_\Omega dv_\ep \cdot \phi = \int_\Omega dv\cdot \phi_\ep=0$ for every
$\psi\in C^1_c(\Lambda^2\Omega)$,  and it follows that $dv_\ep=0$ in $\Omega$. 
Then if $B\in \calC'$, 
\[
\int_{\R^3} d^* B\cdot v = \lim_{\ep\to 0}\int_{\R^3} B \cdot dv_\ep = 0.
\]

{\bf Step 2}. Now for $B\in \calC'$, if $\chi\in C^\infty_c(\R^3\setminus \Omega)$,
then $\chi d^*B\in L^2(\Lambda^1\R^3)$ and $d(\chi d^*B) = 0$ in $\Omega$, so
$\int_{\R^3}\chi |d^*B|^2 = 0$
by \eqref{Cprime1}. Hence  $\supp(d^* B)\subset \bar \Omega$.
Then  \eqref{Cprime1} implies that
for any $v\in L^2(\Lambda^1\Omega)$ such that $dv=0$,
\[
0 \ = \ \int_{\R^3} d^*B\cdot (\chara_\Omega v)  =
\int_{\Omega} d^*B\cdot v  
\]
Thus $(d^*B)|_\Omega \in (\ker d)^\perp = d^*H^1_N(\Lambda^2\Omega)$.

{\bf Step 3}. Now, given $\phi \in d^*H^1_N(\Lambda^2\omega)$, 
let $\psi = (-\Delta)^{-1}(\chara_\Omega\phi)$, and let
$B_\phi :=d\psi$. 

Then the fact that $\phi \in d^*H^1_N$ implies that 
$d^*(\chara_\Omega\phi) = 0$ on $\R^3$. It follows that $d^*\psi = 0$,
and hence that $d^*B_\phi = d^* d\psi = (d^*d + d d^*)\psi = -\Delta\psi = \chara_\Omega\phi$.
In particular $\supp(d^*\psi)\subset \bar \Omega$. 

Finally, to see that $B_\phi\in \calC'$, observe that
$\chara_\Omega\phi\in \cap_{1\le p\le 2}L^p(\Lambda^1\R^3)$, so elliptic regularity and embedding theorems imply $B_\psi \in \cap_{1<p\le 2}W^{1,p} \subset \cap_{3/2<q<6}L^q$, 
$B_\phi\in H^1(\Lambda^2\R^3)\cap d\dot H^1(\Lambda^1\R^3)$,
and in addition \eqref{Creg} holds.
It is clear that $\supp(d^*B_\phi) = \supp (\chara_\Omega\phi)\subset \bar \Omega$,
so $B_\phi\in \calC'$. 

\end{proof}

We need one more easy fact about $\calC'$.

\begin{lemma}If $B_1, B_2\in \calC'$, then there exists $\psi_1\in \cap_{1<p\le 2}\dot W^{2,p}(\Lambda^1\R^3)$ such that $d\psi_1 = B_1$ and
\[
\int_{\R^3} B_1\cdot B_2 \ = \ \int_{\Omega} \psi_1 \cdot d^* B_2.
\]
\label{L.dinv}\end{lemma}

\begin{proof}Let $\psi_1 = (-\Delta)^{-1} d^* B$, so that in view of Remark \ref{Rpsi},
\[
\int_{\R^3} B_1\cdot B_2 \ = 
\int_{\R^3} d\psi _1\cdot B_2 \ = 
\int_{\R^3} \psi _1\cdot d^* B_2 \ = 
\ \int_{\Omega} \psi_1 \cdot d^* B_2
\]
where the integration by parts is easily justified in view of the decay
properties recorded in \eqref{Creg} and Remark \ref{Rpsi}.
\end{proof}

Now we give the

\begin{proof}[proof of Theorem \ref{cor:critfield}]

{\bf Step 1}.
We first assume that $\| B_*\|_* \le \frac 12$. Then, recalling
\eqref{v0eqn}, 
and recalling that $dB_0=0$, we must show that
\[
0=dv_0 
= d d^*B_0 + dA_0 = d d^* B_0 + B_0 + H_{ex} \quad\mbox{ in }\Omega.
\]
Since $B_0$ and $B_*$ minimize $\calE_0$ in $\calC$ and $\calC'$ respectively,
and since $\calC\subset \calC'$,
it is clear that  $B_* = B_0$ if and only if $B_*\in \calC$, which  holds if and only if $\|B_*\|_*\le \frac 12$.
So it suffices to check that
\beq
 dd^* B_* + B_* + H_{ex} = 0\quad\mbox{ in }\Omega.
\label{Bstar.eqn}\eeq
To do this, we take first variations of $\calE_0$ in $\calC'$ to find that $B_*$ satisfies
%By definition, $B_*$ minimizes the functional $\calE_0$, defined in \eqref{dualfunctional},
%in the space $\calC'$, defined in \eqref{constraint'}. Thus
\beq
\int_{\R^3} B_*\cdot B + \int_{\Omega}( d^*B_* + A_{ex})\cdot d^* B = 0
\quad\quad\quad\mbox{for all $B\in \calC'$. }
\label{Bstar.weakEL}\eeq
By Lemma \ref{L.dinv}, we may rewrite this as
\[
\int_\Omega (\psi + d^*B_* + A_{ex}) \cdot d^*B = 0\quad\quad\quad\mbox{for all $B\in \calC'$}.
\]
where $\psi = (-\Delta)^{-1}d^*B_*$, so that $d\psi = B_*$. Then we conclude from Lemma \ref{LCprime} that $(\psi + d^*B_* +A_{ex})|_\Omega\in ((\ker d)^\perp)^\perp = \ker d$,
and hence that $d(\psi + d^*B_* +A_{ex}) = 0$ in $\Omega$, which is \eqref{Bstar.eqn}.

{\bf Step 2}. Now we assume that $dv_0=0$ in $\Omega$.
We will show that in this case, $\calE_0(B_0) = \calE_0(B_*)$. Since $B_*$ is the unique minimizer
of $\calE_0$ in $\calC'$ and $B_0\in \calC\subset \calC'$, this implies that $B_0 = B_*$, and hence that $B_*\in \calC$.

First note that
\beq
\int_\Omega v_0\cdot d^*B = 0
\quad\quad\quad
\mbox{  for any $B\in \calC'$},
\label{kf1}\eeq
since $d^*B\in d^*H^1_N(\Lambda^2\Omega) = (\ker d)^\perp$
by Lemma \ref{LCprime}.
Applying this to $B=B_0$ and recalling that $v_0 =  d^*B_0 + A_0 = d^*B_0 + (A_0-A_{ex}) + A_{ex}$ in $\Omega$, we obtain
\begin{align*}
0 
&= \int_{\Omega} (d^*B_0 + (A_0-A_{ex}) + A_{ex})\cdot d^*B_0\\
&= \int_{\R^3} (d^*B_0 + (A_0-A_{ex}) + A_{ex})\cdot d^*B_0\\
&= \int_{\R^3} |d^*B_0|^2 + |B_0|^2  + A_{ex}\cdot d^*B_0.
\end{align*}
(The integration by parts is easily justified using \eqref{Creg}.) 
Using this to rewrite the definition of $\calE_0$ yields
\beq
\calE_0(B_0) = 
\frac 12 \int_{\Omega} A_{ex}\cdot d^*B_0 + |A_{ex}|^2
\label{Bzeroequip}\eeq

{\bf Step 3}. Next, taking $B_*$ as a test function in \eqref{Bstar.weakEL}, we obtain
\begin{align*}
\int_{\R^3} |B_*|^2  +  \chara_\Omega(|d^*B_*|^2 + A_{ex} \cdot d^*B )
& = 0.\\
\end{align*}
It follows that
\beq
\calE_0(B_*) 
\ = \  \frac 12 \int_{\Omega}( d^*B_*\cdot A_{ex} + |A_{ex}|^2)
\label{Bstarequip}
\eeq

{\bf 3}. 
From \eqref{Bstar.weakEL} we also have
\beq%\begin{align*}
0 = \int_{\R^3} B_* \cdot B_0 + \chara_{\Omega} (d^*B_*+A_{ex})\cdot d^*B _0 = 0
\label{Aex1}\eeq
On the other hand, again using \eqref{kf1}, we compute
\begin{align*}
0
= 
\int_{\Omega} v_0\cdot d^*B_* 
&=
\int_{\Omega} (d^* B_0 + (A_0 - A_{ex}) + A_{ex})\cdot d^*B_* \\
&=
\int_{\R^3} (d^* B_0 + (A_0 - A_{ex}) +  \chara_\Omega A_{ex})\cdot d^*B_* \\
&=
\int_{\R^3} d^*B_0\cdot d^*B_*  +  B_0\cdot B_* + \chara_\Omega  A_{ex}\cdot d^*B_* 
\end{align*}
recalling that $d(A_0-A_{ex}) = B_0$. And by comparing this and \eqref{Aex1}, we 
find that
\[
\int_\Omega A_{ex} \cdot d^* B_* = \int_{\Omega} A_{ex}\cdot d^*B_0.
\]
%\\
%&=
%\int_{\R^3} B_* \cdot B_0 + \chara_{\Omega} (d^*B_*+A_{ex})\cdot (v_0-A_0)\\
%&=
%\int_{\R^3} B_* \cdot B_0 + \chara_{\Omega} (d^*B_*+A_{ex})\cdot (v_0-A_0)\\
%\end{align*}
This, together with \eqref{Bzeroequip} and \eqref{Bstarequip}, shows that $\calE_o(B_*) = \calE_0(B_0)$, completing the proof.

\end{proof}

\subsection{an alternate characterization of the critical applied field.}
\label{S:3.2}

Our next result gives a different characterization of the critical field.

\begin{theorem}
Let $(v_0, A_0)$ minimize $\calF$ in $ L^2(\Omega;\Lambda^1\R^3)\times [A_{ex}+\dot H^1_*]$.

Further, let 
\beq
\calE_1(A) =  \frac 12 \int_{\R^3}\chara_\Omega|PA|^2 +  |dA - H_{ex}|^2 \,dx,
\label{calE.def}\eeq
where $P$ is defined in \eqref{morrey1},
and let $A_1$ minimize $\calE_1$ in $A_{ex}+ H^1(\R^3;\Lambda^1\R^3)$.
Let $\alpha_1\in H^1(\Omega;\Lambda^2\R^3)$ be such that
\beq
d^*\alpha_1 = PA_1, \ d\alpha_1 = 0\quad\mbox{ in }\Omega,\quad\quad\quad \alpha_{1,N} = 0\quad\mbox{ on }\partial \Omega.
\label{alpha0}\eeq
(Such an $\alpha_1$ exists by definition of $P$.) Note that $A_1$ and hence $\alpha_1$ depend on $A_{ex}$.

Then $dv_0 = 0$ if and only if
\beq
\| \alpha_1 \|_{**} := \sup_{|dv|(\Omega) \le 1} \int_\Omega dv\cdot  \alpha_1 \le 1/2.
\label{subcrit}\eeq
Moreover, if $dv_0=0$ then $A_0=A_1$.
\label{thm:GLcritical}\end{theorem}

\subsubsection{Theorem \ref{thm:GLcritical} via a splitting of $\calF$.}

We will give three proofs of this theorem. We first present the most direct proof, which does not
use convex duality at all. 

\begin{proof}[First proof of Theorem \ref{thm:GLcritical}]   
Recall from \eqref{kerd.char} that for $v\in L^2(\Lambda^1\Omega)$,  $dv = 0$ in $\Omega$ 
if and only if $v\in dH^1(\Omega) \oplus \calH_N(\Lambda^1\Omega) = \ker d$, see \eqref{kerd.char}. Define
\begin{align*}
\tilde \calF(v,A) &:= \inf \{\calF(v+\gamma, A) : \gamma \in \ker d \}\\
&\overset{\eqref{morrey}}=
 \frac 12\left[\int_\Omega |dv|+ |P(v-A)|^2 dx \ + \ \int_{\R^3} |dA - H_{ex}|^2 \,dx
\right].
\end{align*}
(Note that  the definition \eqref{calE.def}  of $\calE_1$ can be rewritten  $\calE_1(A) = \tilde \calF(0, A)$.)
It is clear that 
\[
(v_0, A_0) \mbox{ minimizes }\tilde \calF\quad \iff \quad
(v_0 +\gamma, A_0)\mbox{ minimizes }\calF\mbox{ for some $\gamma\in \ker d$}.
\]
%\[
%\inf \calF \le \tilde \calF(v,a) \le \calF(v,a),\quad  \mbox{ with equality if $(v,a)$ minimizes $\calF$.}
%$\calF(\cdot, \cdot)$.}
%\]
%and 
Since we are interested here in $dv_0$,  we may consider $\tilde \calF$ instead of $\calF$.
We rewrite
\[
\tilde \calF(v,A) = \calE_1(A) +  \frac 12 \int_\Omega |dv|+|Pv |^2 - 2Pv \cdot  P A  \,dx .
\]
Since $A_1$ minimizes $\calE_1$, 
\beq
\int_{\R^3} \chara_\Omega PA_1\cdot P B + (dA_1-H_{ex})\cdot dB \ dx = 0
\label{wkel}\eeq
for all  $B\in   \dot H^1(\R^3;\Lambda^1\R^3)$, so that
\[
\calE_1(A_1+B) = \calE_1(A_1) +  \frac 12 \int_{\R^3} \chara_\Omega |PB|^2 + |dB|^2 \ dx
\]
for  $B$ as above.
Given any $A$, let us write $A = A_1+B$. Then 
\begin{align*}
\tilde \calF(v,A_1+B) 
&=
\calE_1(A_1) + \frac 12  \int_\Omega |PB|^2 \, dx \ + \ \int_{\R^3} |dB|^2 \ dx
%\int_{\R^3} \chara_\Omega A_1\cdot b + (dA_1-H_{ex})\cdot db \ dx
\\
&
\quad\quad\quad
+ \frac 12 \int_\Omega |dv|+ |Pv |^2 - 2 Pv \cdot( P A_1 + PB) \,dx \\
&=
\frac 12 \int_\Omega |P(B-v)|^2 +\frac 12 \int_{\R^3} |dB|^2  + \int_\Omega
\frac 12 |dv| - Pv \cdot PA_1. 
\end{align*}
For $\alpha_1$ as in the statement of the theorem,
\beq
\int_\Omega
 Pv \cdot PA_1
 \ = \ 
\int_\Omega  P v\cdot d^*\alpha_1 \ dx 
\ = \ 
\int_\Omega  dv \cdot \alpha_1\  dx 
\label{thm3pf1.1}\eeq
where the boundary terms arising from integration by parts have vanished due to the
fact that $\alpha_{1,N}=0$.
Thus
\beq
\tilde \calF(v,A) = \calE_1(A_1) + \frac 12 \int_{\R^3}|dB|^2 + \chara_\Omega |P(v-B)|^2 
+ \int_{\Omega} ( \frac 12 |dv| - dv\cdot \alpha_1).
\label{split}\eeq
If condition \eqref{subcrit} holds, then $ \int_{\Omega}( \frac 12 |dv|- dv\cdot \alpha_1) \ge 0$
for all $v\in L^2(\Omega)$, and thus $\tilde \calF(v,A) \ge \calE_1(A_1)$ for all $(v,A)$.
Moreover, if $(v_0, A_0) = (v_0, A_1+B_0)$ attains this minimum, then 
$\frac 12 \int_{\R^3}|dB_0|^2 + \chara_\Omega |P(v_0-B_0)|^2 =0$, and this implies 
that $dv_0=0$.

And if  \eqref{subcrit} fails, then there exists some $v_1$ such that
$ \int_{\Omega}- dv_1\cdot \alpha_1 + \frac 12 |dv_1| < 0$, and then it is clear that
$\tilde \calF(\lambda v_1, A_1) < \calE_1(A_1) = \tilde \calF(0, A_1)$ for all sufficiently small $\lambda>0$.
Thus $\tilde \calF(v_0, A_0) < \calE_1(A_1)$ for any minimizing $(v_0, A_0)$, and then
\eqref{split} implies that $dv_0\ne 0$.

Finally, it $dv_0=0$ then it is clear from \eqref{split} that $\tilde \calF(0,A_1) = \calE_1(A_1) = 
\min \calF$, and and hence that $A_0=A_1$.
\end{proof}

\subsubsection{Theorem \ref{thm:GLcritical} via partial convex duality}

We next prove Theorem \ref{thm:GLcritical} 
by a duality computation that differs slightly from the one
used in the proof of Theorem \ref{thm:GLmain}. 
The result of this computation is summarized in the following

\begin{lemma} 
Let
\beq
N := \left\{ \zeta \in L^2(\Omega) :  ( \zeta,  \xi)_{L^2(\Omega)} \le  \frac 12\int |d\xi| \quad\mbox{ for all }\xi\in L^2(\Omega) \right\}.
\label{N.def}\eeq
and define
\beq\label{eq:Fddag}
\calF^\ddagger(A) :=
 \frac 12 \int_{\R^3}( |d(A-A_{ex})|^2 + \chara_\Omega |A|^2 ) \ dx - \frac 12 \dist_{L^2(\Omega)}^2(A,N)
\eeq
Then $(v_0,A_0)$ minimizes $\calF$ in
$L^2(\Omega;\Lambda^1\R^3)\times [ A_{ex}+ \dot H^1_*]$
if and only if 

{\bf 1}. $A_0$ minimizes $\calF^\ddagger$ in $[ A_{ex}+ \dot H^1_*]$, {\em and}

{\bf 2}. 
%$v_0$ is such that $A_0|_\Omega - v_0$ is the closest point to $A_0$ in $N$ .
$A_0|_\Omega - v_0\in N$, and $\| A_0 - v_0\|_{L^2(\Omega)} = \dist_{L^2(\Omega)}(A_0,N)$.
\label{L.dualb}\end{lemma}

It is clear from the definition that $N\subset (\ker d)^\perp = \mbox{Image}(P)$,  and it follows that
%we can rewrite $\calF^\ddagger$ as
\beq\label{eq:Fddagbis}
\begin{aligned}
\calF^\ddagger(A) &=\frac 12 \int_{\R^3}( |d(A-A_{ex})|^2 + \chara_\Omega |PA|^2 ) \ dx
 - \frac 12 \dist_{L^2(\Omega)}^2(PA,N)\\
&= \calE_1(A)  - \frac 12 \dist_{L^2(\Omega)}^2(PA,N).
\end{aligned}
\eeq

\begin{proof}
We will compute the convex dual of $\calF$ with respect to the ``$v$'' variable only, treating
$A$ as a parameter.
Thus, let $\xi = A|_\Omega - v$, and write $\tilde F(\xi; A) = \calF(v, A)$,
so that
\[
\tilde F(\xi ;A) = \frac 12 \int_{\Omega} |\xi|^2 + |d(A-\xi)| + \frac 12 \int_{\R^3} |d(A-A_{ex})|^2.
\]
Let 
\[
\tilde I(\xi; A) := \frac 12\int_{\Omega} |d(A-\xi)| + c_A,\quad\quad\quad c_A :=  \frac 12 \int_{\R^3} |d(A-A_{ex})|^2.
\]
Then $\tilde F(\xi; A) = \tilde I(\xi; A) + \frac 12 \|\xi\|_2^2$.

Next, let
\[
\tilde I^*(\xi^* ; A) := \sup_{\xi\in L^2}
\left\{ ( \xi^*, \xi)_{L^2(\Omega)} -\tilde  I(\xi ;A) \right\}.
\]
A short computation like that in the proof of Theorem \ref{thm:GLmain} shows that
\beq
\tilde I^*(\xi^*; A)
= 
\left\{ \begin{array}{ll} 
(\xi^*, A)- c_A\quad
&\mbox{ if }\xi^*\in N\\
+\infty &\mbox{ if not}
\end{array}\right.
\label{tildeIstar}\eeq
for $N$ as defined in \eqref{N.def}.
Now let 
\begin{align*}
\tilde F^\ddagger(\xi^*, A) 
&:= 
\tilde I^*(-\xi^*; A) + \frac 12 \| \xi \|^2\\
&= 
\left\{ \begin{array}{ll} 
(-\xi^*, A)+ \frac 12 \|\xi\|^2- c_A\quad
&\mbox{ if }\xi^*\in N\\
+\infty &\mbox{ if not}
\end{array}\right.\\
&=
\left\{ \begin{array}{ll} 
\frac 12 \| \xi^* - A\|^2 - \frac 12 \int_{\R^3}( |d(A-A_{ex})|^2 + \chara_\Omega |A|^2 ) \ dx 
&\mbox{ if }\xi^*\in N\\
+\infty &\mbox{ if not.}
\end{array}\right. 
\end{align*}
Then it is clear  that 
\[
- \inf_{\xi^*} \tilde F^\ddagger(\xi^*; A) = \calF^\ddagger(A) 
\]
as defined above, and that the infimum is 
attained by a unique $\xi$, the closest point to $A$ in the (closed convex) set $N$.
Recall from Lemma \ref{lem:dual} that
$\xi$ minimizes $\tilde F(\cdot; A)$ if and only if it minimizes $\tilde F^\ddagger(\cdot;A)$,
and moreover that $\min_\xi\tilde F(\cdot; A) = - \min_\xi\tilde F^\ddagger(\cdot; A)$.
Thus
\[
\min_{A,v} \calF(v;A) 
= \min_A \min_v \calF(v; A) = \min_A( -  \min_\xi \tilde F(\xi, A)) \\
= \min_A \calF^\ddagger(A),
\]
and  $(v_0,A_0)$ minimizes $\calF$ if and only if $A_0$ minimizes $\calF^\ddagger$ and
$v_0 = A_0|_\Omega - \xi_0$, where $\xi_0$ is the closest point in $N$ to $A$.

\end{proof}

%\begin{remark}
%A slight variation of the above argument shows that for any $A$,
%\[
%\inf_{\xi = v-A\in L^2(\Omega)} G(v,A) = \frac 12 \int_{\R^3} |d(A-A_{ex})|^2 dx +  \frac 12 \Psi(A),
%%\]
%where 
% $\Phi(A)  :=  \|A\|^2_{L^2(\Omega)} - \mbox{dist}(A, N)^2$. 
% \texttt{It should be easy to check that
% $\Phi$ is convex. It also has linear growth for $A$ large. One could go on to 
%dualize $\frac 12 \int_{\R^3} |d(A-A_{ex})|^2 dx +  \frac 12 \Phi(A)$ and get yet another  obstacle-type problem....
%But there are many ways of dualizing, and I don't know if this one has anything to recommend it. In fact I'm
%not sure this remark adds anything. }
%\end{remark}

Now we use  Lemma \ref{L.dualb} to give a

\begin{proof}[second  proof of Theorem \ref{thm:GLcritical}]
Fix $(v_0, A_0)$ minimizing $\calF$, and $A_1$ minimizing $\calE_0$.
By Lemma \ref{L.dualb}, $A_0$ minimizes $\calF^\ddagger$.

We write $\xi_0 = A_0|_\Omega - v_0$ as above, so that $\xi_0$
is the closest point to $A_0$ in $N$.
From the definition \eqref{morrey1} of $P$ we know that
%\eqref{morrey1} implies that 
$dPA_0 = dA_0$, so that
\[
dv_0 = 0 
\iff 
d (PA_0 - \xi_0) = 0 \mbox{ in }\calD'(\Omega ).
\]
Also, $PA_0 - \xi_0 \in (\ker d)^\perp = d^* W^{1,2}_N(\Lambda^2 \Omega)$,
since the definitions of $N$ and $P$ imply that $N\subset (\ker d)^\perp$
and Image$(P)=(\ker d)^\perp$. Then \eqref{kerd.char} implies that
$d(PA_0-\xi_0) = 0$ in $\calD'$ if and only if $PA_0-\xi_0 \in d^* W^{1,2}_N(\Lambda^2 \Omega)\cap
\left(dH^1(\Omega)\oplus \calH(\Lambda^1\Omega)\right) = \{0\}$.
In other words, $dv_0=0$ if and only if $PA_0 = \xi_0$.
But since $\xi_0$ is the closest point in $N$ to $A_0$, and hence to $PA_0$,
we conclude that
\beq 
dv_0 = 0 \ \ \iff \ \ 
PA_0\in N .%, \mbox{ where $A_0$ minimizes $\calF^\ddagger$ in $H^1_{*,A_{ex}}$.} \ 
%d^*(dA_0- H_{ex}) + \chara_\Omega PA_0 = 0 \mbox{ in }\R^3 .   
\label{thm3pf2.1}\eeq

Next, note that $A\mapsto \calF^\ddagger(A_{ex}+ A)$ is strictly convex in
$H^1_*$, so that  the minimizers $A_0$ of $\calF^\ddagger$ and $A_1$ of $\calE_1$ 
are unique.
Also, \eqref{eq:Fddagbis} implies that if $PA\in N$ and $A'\in H^1_*$, then
\[
\calF^\ddagger(A + A') \le  \calE_1(A + A') \le \calF^\ddagger(A) + C\| A'\|_{L^2(\Omega)}^2 \le
\calF^\ddagger(A) + C\| A'\|_{H_*}^2.
\]
Thus any critical point $A$ of $\calE_1$ such that  $PA\in N$ must also be a critical point of $\calF^\ddagger$, and conversely. 
\beq
\mbox{ $PA_0\in N$ if and only if $PA_1\in N$.}
\label{thm3pf2.2}\eeq
It follows along the same lines that if $dv_0=0$ then $A_0=A_1$.
Finally, recalling the definitions \eqref{alpha0} of $\alpha_1$ and \eqref{N.def} of $N$,
and integrating by parts as in \eqref{thm3pf1.1}, we conclude that
\[
PA_1\in N  \ \ \iff \ \  \| \alpha_1\|_\star \le \frac 12.
\]
By combining this with \eqref{thm3pf2.1} and \eqref{thm3pf2.2}, we conclude this
proof of Theorem \ref{thm:GLcritical}.
\end{proof}

\subsubsection{Equivalence of Theorem \ref{thm:GLcritical} and Theorem \ref{cor:critfield}.}

In Theorem \ref{cor:critfield} and Theorem \ref{thm:GLcritical}, we have derived two necessarily equivalent but rather different-looking necessary and sufficient conditions for
the vorticity $dv_0$ of a minimizing pair $(v_0,A_0)$ to vanish.
In this section we  elucidate the connection between the auxiliary functions
$B_*$, defined in Theorem \ref{cor:critfield}, and $\alpha_1$, defined in \eqref{subcrit}.

\begin{proposition}  If $dv_0=0$ in $\Omega$, then $PA_1 = d^*\alpha_1 = (d^*B_*)|_\Omega$, 
and  $\| \alpha_1\|_{**} = \| B_*\|_*$.
\end{proposition}

This can be seen as a third proof of Theorem \ref{thm:GLcritical}.

\begin{proof}
If $dv_0=0$, then $Pv_0=0$, and we have seen that $A_0= A_1$ and $B_0 = B_*$.
As a result,   
\[
(d^*B_*)|_\Omega= (d^*B_0)|_\Omega = P(d^*B_0) =P(A_0-v_0)
= PA_0 = PA_1 =d^*\alpha_1
\]
by  Theorem \ref{cor:critfield}, Lemma \ref{LCprime},  and \eqref{v0eqn}.
Thus for every $v\in H^1(\Lambda^2\R^3)$,
\[
\int_{\R^3} dv \cdot B_* = \int_{\R^3} v \cdot d^*B = \int_{\Omega} v \cdot d^*B
= \int_{\Omega} v \cdot d^*\alpha_1 = \int_{\Omega}dv\cdot \alpha_1.
\]
Now the conclusion follows from the definitions of  the norms $\| \cdot \|_*$ and 
$\| \cdot \|_{**}$, see \eqref{weirdnorm} and \eqref{subcrit}.
\end{proof}

\section{vortex density in 3d Bose-Einstein Condensates}\label{sect:GP}

%{\bf add remark about higher rotations and equidistribution of vortex lines}

In this section we use the results of \cite{BJOS1} to prove
convergence as $\e\to 0$ of Gross-Pitaevsky functional $\calG_\ep$, defined in \eqref{Gep.def}
to the limiting $\calG$, defined in \eqref{G0.def}. We also establish some results describing
minimizers of $\calG$.

\subsection{$\Gamma$-convergence}\label{S:becGlim}

Our first theorem makes precise the sense in which $\calG$ is a limiting functional associated to the
sequence of functionals  $(\calG_\e)_{\e\in (0,1]}$. The statement of the result
uses some notation that is introduced in Section \ref{sect:GPintro}.

\begin{theorem}\label{GP.Gammalim}
Assume that 
$\Phi_\e = \logeps \Phi$ for $\Phi\in L^4_{loc}(\Lambda^1\R^3)$
and that $|\Phi(x)|^2 \le C (a(x) +1)$ for all $x\in \R^3$.

{\rm (i) compactness}: Assume that $(u_\e)_{\e\in (0,1]} \subset H^1_{a,m}$ and that there exists some $C>0$ such that
\beq\label{Glim.ch}
\calG_\e(u_\e) \le  C \logeps^2
\quad\quad\mbox{ for all }\e\in (0,1/2]
\eeq
Then 
%\[
%|u_\e|^2 \to \rho\quad\mbox{ in }L^4(\R^3)
%\] and 
there exists $j\in L^{4/3}(\Lambda^1\R^3)$, supported in $\bar \Omega$, such that if we define $v = \frac 1\rho j|_\Omega$ and pass to a subsequence if necessary, we have
\beq
\frac{ju_\e}{\logeps} \rightharpoonup j = \rho v \mbox{ weakly in }L^{4/3}(\R^3), \ 
\mbox{ and }\mbox{ $v\in L^2_\rho(\Lambda^1\Omega)$ with }\int_\Omega\rho |dv|<\infty.
%\mbox{ for  certain }p>1
\label{GP.compact}\eeq

{\rm (ii): lower bound inequality}: There exists a sequence of numbers $(\kappa_\e)$ such that
if we assume above hypotheses and \eqref{GP.compact}, then
\beq\label{GP.lbd}
\liminf_{\e\to 0}  \logeps^{-2}\left( \calG_\e(u_e) - \kappa_\ep \right) \ge \calG(v).
\eeq

{\rm (iii) upper bound inequality} Given any $v\in L^2_\rho(\Lambda^1\Omega)$ such that $dv$ is a measure on $\Omega $ with $\int_\Omega \rho|dv| <\infty$,
there exists a sequence $u_\e\in H^1_{a,m}$ such that
 \eqref{GP.compact} holds and
$\lim_{\e\to 0} \logeps^{-2}\left( \calG_\e(u_e) - \kappa_\ep \right) = \calG(v)$.

\end{theorem}

%\begin{remark}\label{rem:thm4} Assumption $\Phi\in C^{2,\alpha}$ is just technical, in order to use Proposition \ref{prop:montero}, and can be relaxed to $\Phi\in L^4(\Lambda^1\Omega)$, which ensures continuity of the term $\frac{1}{\logeps}\int_\Omega ju_\eps\cdot\Phi$ with respect to weak $L^{4/3}$ convergence of  $\frac{1}{\logeps}ju_\eps$.
%\end{remark}

The theorem states that the functionals $ \logeps^{-2}( \calG_\e(\, \cdot \,) - \kappa_\ep )$ converge to 
$\calG$ in the sense of $\Gamma$-convergence, with respect to the convergence
\eqref{GP.compact}.
As remarked in the introduction, Proposition \ref{prop:GPcompact} is a direct corollary of
Theorem \ref{GP.Gammalim} and basic properties of $\Gamma$-convergence.

\begin{remark}\label{rem:higher} In fact we prove a more general result
than Theorem \ref{GP.Gammalim}, since we also allow higher 
rotations $\Phi_\ep=\sqrt{g_\ep}\Phi$, with $\logeps^2\ll g_\ep\ll\ep^{-2}$. In fact 
we show that for such $\Phi_\ep$, if $\calG_\ep(u_\ep)\le Cg_\ep$, 
then after passing to a subsequence,
$\frac{ju_\ep}{\sqrt{g_\ep}}\rightharpoonup j$ weakly in $L^{4/3}(\Lambda^1\R^3)$, with $j = \chara_\Omega\rho v$ for some  $v\in L^2_\rho(\Lambda^1\Omega)$, and
\beq\label{GP.higher}
g_\ep^{-1}\left(\calG_\ep(\cdot) -\kappa_\ep\right)\xrightarrow{\Gamma}\tilde\calG(\cdot)\, , 
\qquad\mbox{where } 
\tilde\calG(v)=\int_{\Omega}\rho\left(\frac{v^2}{2}-\Phi\cdot v\right)\, .
\eeq
As noted in Remark \ref{rem:higher1} in the Introduction,
it easily follows that for  rotations around the $z$ axis of order $\logeps \ll \sqrt{g_\ep} \ll \ep^{-1}$, ground states
exhibit an asymptotically uniform distribution of vertical vortex lines, generalizing  2d results of 
\cite{CorrYng2008}.
\end{remark}

The proofs rely at certain points on Theorem 2 in \cite{BJOS1}. 
%Theorem \ref{GP.Gammalim} is in some sense a corollary of Theorem 2 in \cite{BJOS1}, whichmore or less corresponds to Theorem \ref{GP.Gammalim} in the case (not covered here) in which $a = 0$ in $\Omega$ and $a = +\infty$ on $\R^3\setminus \Omega$,  $m = |\Omega|$, and $\Phi_\e \equiv 0$. 

\begin{proof} [Proof of Theorem \ref{GP.Gammalim} and Remark \ref{rem:higher}]
Let $\calG_\ep(u_\ep)\le Cg_\ep$, for $\logeps^2\le g_\ep\ll\ep^{-2}$, and let $\Phi_\ep=\sqrt{g_\ep}\Phi$.

{\bf Step 1}. First we control the potentially negative term in $\calG_\e(u_\e)$. To do this, 
recall our assumption that $|\Phi|^2 \le C(a + 1)$. Since
$|ju| \le |u|\ |du|$, it follows that
\begin{align*}
|\Phi_\e \cdot ju_\e |
\ \le \  
\frac 12
g_\ep |\Phi|^2 |u|^2 + \frac 12|du|^2
\ \le  \  Cg_\ep (a+1) |u|^2 + \frac 12 |du|^2.
\end{align*}
But \eqref{rho.def} implies that $a(x) = w(x) - \rho(x) + \lambda \le w(x) +\lambda$, so it follows that
\[
|\Phi_\e \cdot ju_\e |
\le Cg_\ep( w + \lambda +1)|u_\e|^2 +  \frac 12 |du_\e|^2.
\]
Integrating this over $\R^3$ and recalling that $\|u_\e\|_2^2 = m$, we obtain
\[
\int_{\R^3} |\Phi_\e \cdot ju_\e | \le C \calG_\e(u_\e) +  C g_\ep( \lambda+1) \int_{\R^3}|u_\e|^2 \le
C g_\ep.
\]
It follows from this that
\beq
\int_{\R^3} \frac 12 |du_\e|^2 + \frac 1{4\ep^2}(\rho-|u_\e|^2)^2 + \frac {w}{2\e^2}|u_\e|^2= \calG_\e(u_\e) + \int_{\R^3} \Phi_\e \cdot ju_\e  \le C g_\ep.
\label{GPest0}\eeq
In particular $|u_\e|^2 \to \rho$ in $L^2(\R^3)$.

{\bf Step 2}. Next,
\[
\| u_\e\|_{L^4(\R^3)}^4 = \| \ |u_\e|^2 \|_{L^2}^2 \le  C ( \| \ |u_\e|^2- \rho  \|_{L^2}^2 +  \|  \rho  \|_{L^2}^2 )
\le C + \e^2 \calG_\e(u_\e) \le C,
\]
and Step 1 implies that $\| du_\e\|_{L^2} \le C \sqrt{g_\ep}$. Since $\|ju_\e\|_{L^{4/3}}\le \| u_\e\|_{L^4}\|du_\e\|_{L^2}$, we conclude that
$\{ \frac{1}{\sqrt{g_\ep}} ju_\e \}$ is uniformly bounded in $L^{4/3}(\R^3)$, and it follows that there exists some $j\in L^{4/3}(\R^3)$ such that 
\beq\label{GPcompact.a}
\mbox{$\frac{ ju_\e}{\sqrt{g_\ep}}\rightharpoonup j$ weakly in $L^{4/3}$ along some subsequence.}
\eeq

{\bf Step 3}.
Now let $f_\e$ denote the minimizer in $H^1_a(\R^3;\R)$ of  $\calG_\e(\cdot)$, where $H^1_a(\R^3;\R)$ is defined by analogy with $H^1_a(\R^3;\C)$, see \eqref{H1a.def}.

Note that when $f$ is real-valued, $jf= 0$, so the forcing term $\Phi\cdot jf$ vanishes
on $H^1_a(\R^3;\R)$. It is standard that $f_\e$ does not vanish, and we will assume that $f_\e>0$.
Then for any $u\in H^1_{a,m}$ we may define $U := u/f_\e$, and
it is known that 
\beq
\calG_\e(u) = \calG_\e(f_\e U) = \calG_\e(f_\e) \ + \ H_\e(U)  \ + \   \int_{\R^3}  f_\e^2\Phi_\e\cdot jU \
,
\label{GP.split}\eeq
where $H_\e(U) = H_\e(U; \R^3)$, and for a measurable subset $A\subset\R^3$ we write
\beq
H_\e(U; A) := \int_{A} \frac {f_\e^2}2 |dU|^2  + \frac {f_\e^4}{4\e^2}(|U|^2-1)^2 \ dx.
\label{Hep.def}\eeq
See  \cite{Montero2008} for a proof in exactly\footnote{The Gross-Pitaevsky integrand is written in a slightly different way in  in \cite{Montero2008}, but this is purely a cosmetic difference.}  the situation we consider here, following ideas that originated in \cite{LM99} and have been used extensively in the literature on Bose-Einstein condensates.  It is also known that $f_\e^2\to \rho$ uniformly in $\R^3$, see again \cite{Montero2008}.

{\bf Step 4}. Now let $\Omega'$ denote an subset of $\Omega$ such that $\Omega'\subset\subset\Omega$, so that $\rho\ge 2 c'$ in $\Omega'$ for some $c'$, and hence $f_\e^2\ge c'$ for all sufficiently small $\e$. 
 Let $U_\e = u_\e/ f_\e$, and note that
Step 1 implies that 
$H_\e(U_\e)\le Cg_\ep$. Thus the functional
\[
\tilde H_\ep(U_\ep)=\int_{\Omega'} \frac 12|dU_\e|^2 + \frac {1}{\ep^2}W(u) \ \le \ C'g_\ep ,\qquad\mbox{where }W(U)= \frac{c'}4(|U|^2-1)^2,
\]
verifies in $\Omega'$ hypothesis $(H_q)$ of Theorem 2  in \cite{BJOS1}, for $q=4$. Hence Theorem 2 of \cite{BJOS1} (or  arguments such as those in Steps 1 and 2 above) imply  that there exists some $v'$ in $L^2(\Lambda^1 \Omega')$
such that, after passing to a further subsequence if necessary,
$
\frac{jU_\e}{\sqrt{g_\ep}} \rightharpoonup v'\quad\mbox{ weakly in } L^{4/3}(\Lambda^1 \Omega').
$
Since $u_\e = f_\ep U_\ep$, one easily checks that $ju_\ep = f_\ep^2 \, jU_\ep$, and then it follows
from \eqref{GPcompact.a} and the uniform convergence $f_\ep^2\to \rho$ that $j = \rho v'$ in
$\Omega'$, and hence that $v' = j/\rho =: v$ in $\Omega'$, independent of $\Omega'$. It also follows that the chosen subsequence is independent of $\Omega'$. 
%We now write $\e' := \e / c'$, so that $f_\ep^2/\e^2 \ge 1/\ep'^2$ in $\Omega'$, and thus
Let moreover
\[
\mu_\ep := \frac 1{g_\ep}\left( \frac 12|dU_\e|^2 + \frac {1}{\ep^2} W(u)\right)\, dx,
\]
be the energy density of $\tilde H_\ep$,
and notice also that $\int_{\Omega'}\mu_\ep$ is uniformly bounded, so that after passing to a subsequence, we may assume that there exist a measure $\mu_0$ 
$\Omega'$ such that $\mu_\ep\rightharpoonup \mu_0$ 
weakly as measures in $\Omega'$. It then follows from Theorem 2  and Remark 4 in \cite{BJOS1},  that, in the case $g_\ep\le C\logeps^2$,
$\mu_0 \ge \frac 12( |v|^2\,dx + |dv|)$, in the sense that $|dv|$ is a Radon measure, and 
\[
\mu_0(U) \ge \frac 12  \int_U( |v|^2\, dx + |dv|)
\]
for every open $U\subset \Omega'$, while for $\logeps^2\ll g_\ep\ll\ep^{-2}$ we have
$\mu_0\ge \frac 12 |v|^2\,dx$. In either case we deduce (using basic facts about weak convergence of measures) 
that
\beq
\liminf_{\e\to 0} \frac 1{g_\ep}H_\e(U_\e; \Omega') = 
\liminf_{\e\to 0} \int_{\Omega'}f_\ep^2 \mu_\ep \ge
\int_{\Omega'} \rho \mu_0\, , \label{GP.interior}
\eeq
which yields 
\beq\label{GP.int1}\liminf_{\e\to 0} \frac {1}{g_\ep}H_\e(U_\e; \Omega')\ge \frac{1}{2}\int_{\Omega'}\rho( |v|^2\, dx + |dv|)\qquad\mbox{if }g_\ep\le C\logeps^2,
\eeq
 and
\beq\label{GP.int2}\liminf_{\e\to 0} \frac 1{g_\ep}H_\e(U_\e; \Omega')\ge \frac{1}{2}\int_{\Omega'}\rho|v|^2\, dx\qquad\mbox{if }\logeps^2\ll g_\ep\ll\ep^{-2}.
\eeq
Since this holds for all $\Omega'\subset \Omega$, it follows in particular that 
$v\in L^2_\rho(\Lambda^1\Omega)$ and (in case $g_\ep\le C\logeps^2$) that $dv$ is a measure on all of $\Omega$ with
$\int_\Omega \rho|dv|<\infty$,
nearly completing the proof of \eqref{GP.compact}. (We still need  however to prove that
$j$ is supported in $\bar \Omega$.)

{\bf Step 5}. 
We next claim that
\beq\label{GP.conv1}
\frac 1{g_\ep}\int_{\R^3} f_\ep^2 \Phi_\ep \cdot jU_\ep  \to \int_{\R^3} \rho \,\Phi \cdot v.
\eeq
To prove \eqref{GP.conv1}, since $\Phi_\ep = \sqrt{g_\ep} \Phi$ and $\frac{1}{\sqrt{g_\ep}}ju_\e = \frac{1}{\sqrt{g_\ep}}f_\ep^2 \   jU_\ep \rightharpoonup
j = \rho v$ weakly in $L^{4/3 }(\Omega)$, it is clear that
\[
\frac 1{g_\ep}\int_{\Omega} f_\ep^2 \Phi_\ep \cdot jU_\ep \to \int_{\Omega} \rho \,\Phi \cdot v,
\]
and we only need to show that 
\[
\int_{\R^3\setminus \Omega} \Phi \cdot \frac{ ju_\ep}{\sqrt{g_\ep}} \to 0\, .
\]
Since $\rho = 0$ outside $\Omega$, in this set we have
\[
 |j(u_\ep)| \le |u_\ep|\ |du_\ep|  \le \frac{1}{4\ep} |u_\e |^4+ \frac {3\ep^{1/3}}{4} |du_\ep|^{4/3}
 =\frac {1}{4 \ep} (|u_\e |^2-\rho)^2 +\frac {3\ep^{1/3}}{4} |du_\ep|^{4/3},
\]
whence, for any compact $K\subset\R^3$, we see from \eqref{GPest0} that
\[  
\frac{1}{\sqrt{g_\ep}}\int_{K\setminus\Omega}|ju_\ep|
%\le \frac{C}{\sqrt{g_\ep}}\left(\ep g_\ep +\ep^{1/3}g_\ep ^{2/3}\right)
\le C\left( \ep \sqrt{g_\ep}+(\ep\sqrt{g_\ep})^{1/3}\right).
\]
Thus $\frac{ju_\ep}{\sqrt{g_\ep}} \to 0$ in $L^1(\Lambda^1 (K\setminus\Omega))$ for any compact $K$. This implies that $j=0$ outside $\Omega$, so that the identity $j=\rho v$ holds in all of $\R^3$, finally completing the proof of \eqref{GP.compact}, and it also implies that
\[
\frac{1}{g_\ep}\int_{K\setminus \Omega} \Phi_\ep \cdot  ju_\ep=\int_{K\setminus \Omega} \Phi \cdot \frac{ ju_\ep}{\sqrt{g_\ep}} \to 0
\]
for $K$ compact.
Next, due to \eqref{a.coercive}, \eqref{rho.def} and the assumption that
$|\Phi|^2 \le C (a+1)$, we can find a compact $K$ such that 
$|\Phi|^2 \le C w$ outside of $K$, so that (arguing as in Step 1)
\[
|\Phi \cdot ju_\e|
\le \frac \ep 2 |du|^2 + \frac w{2\ep} \,  |u|^2 \mbox{\ \ \ outside of $K$}.
\]
It follows from this and Step 1 that 
\[
\frac 1{g_\ep} \int_{\R^3\setminus K}
|\Phi_\e \cdot ju_\e|
 \ \le \ 
C \e.
\]
By combining these inequalities, we obtain the claim \eqref{GP.conv1}.

{\bf Step 6}. We now complete the proof of the lower bound inequality.
Note that, by combining \eqref{GP.conv1} with \eqref{GP.int1} and recalling
\eqref{GP.split}, we find that, in case $g_\ep\le C\logeps^2$,
\[
\liminf_{\ep\to 0} \frac 1{\logeps^2}(\calG_\ep(u_\ep)-\calG_\e(f_\e)) \ge
\frac 12\int_{\Omega'} \rho ( |v|^2 + |dv|) -\int_\Omega \rho \Phi \cdot v.
\]
for any open $\Omega'$ compactly contained in $\Omega$. Taking the supremum over all
such $\Omega'$, we obtain \eqref{GP.lbd} with $\kappa_\e = \calG_\e(f_\e)$.
Analogously, in case $\logeps^2\ll g_\ep\ll \ep^{-2}$, using \eqref{GP.int2} in place of
\eqref{GP.int1} we obtain the lower bound part in \eqref{GP.higher}
\[
\liminf_{\ep\to 0} \frac 1{g_\ep}\left(\calG_\ep(u_\ep)-\calG_\e(f_\e)\right) \ge
\int_{\Omega} \rho \left( \frac{|v|^2}{2}-\Phi \cdot v \right) \, .
\]

{\bf Step 7}. Let us prove the upper bound inequality in case $g_\ep\le C\logeps^2$. The proof in the case $\logeps^2\ll g_\ep\ll \ep^{-2}$ follows the same lines and hence is omitted. We will use the following lemma.

\begin{lemma}
Suppose that $v\in L^2_\rho(\Lambda^1\Omega)$ and  that $dv$ is a locally finite 
measure with $\int_\Omega \rho|dv| <\infty$. 
Then for  every $\delta>0$, there  exists $v_\delta\in  C^\infty_c(\Lambda^1\R^3)$ 
such that
\beq\label{GP.dense}
\int_\Omega \rho |v_\delta - v|^2  \ < \  \delta,
\quad\quad\quad
\int_\Omega \rho |dv_\delta| 
\ \le  \   \int_\Omega \rho  |dv| +\delta
\eeq
\label{lem:GPdense}\end{lemma}

The proof is given at the end of this section. Now we use the lemma to complete the proof of the theorem. 

Fix $v$ and $v_\delta$ as in the statement of the Lemma. 

It is proved in \cite{Montero2008}, Lemma  B.1,
that the (positive) function $f_\ep$ appearing in the decomposition \eqref{GP.split} satisfies 
$\| f_\e\|_{L^\infty(\R^3)} \le c_0$, for $c_0$ independent of $\e\in (0,1]$,
and moreover there exists $R>0$ such that $\Omega\subset\subset B_R$, and $0< f_\e(x) \le C e^{- R/\e^{2/3}}$ whenever $|x|\ge R$.

Now for every $\delta>0$, it follows from Theorem 2 and Remark 4 in \cite{BJOS1} that there exists a sequence $U^\delta_\e\in H^1(B_{R+1};\C)$ such that  $\frac 1\logeps jU^\delta_\e \rightharpoonup v_\delta$ weakly in $L^{4/3}(B_{R+1})$,  and
\beq
\frac 1\logeps\left( \frac 12 |d U^\delta_\ep|^2 + \frac 1{4\ep^2}(|U^\delta_\ep|^2-1)^2 \right)
\rightharpoonup \frac 12(|v_\delta|^2 + |dv_\delta|)
\label{vepdelta}\eeq
weakly as measures in $B_{R+1}$.
Let $\e' := c_0\e$, and  
let $u_\e^\delta := f_\e \chi U^\delta_{\e'}$, where $\chi\in C^\infty_c(B_{R+1})$ is a function
such that $\chi\equiv 1$ on $B_R$ and $|d\chi|\le C$. Also,  set $u^\delta_\e := 0$ on $\R^3\setminus B_{R+1}$.
Then as in \eqref{GP.split}, and recalling that $\Phi_\e = \logeps\Phi$, we have
\[
\frac 1{\logeps^2}(\calG_\e(u^\delta_\e) - \calG_\e(f_\e))  \ = \  
\frac 1{\logeps^2}H_\e( \chi U^\delta_{\e'}) + \frac 1 {\logeps^2}\int_{\R^3} \chi  f_\e^2  \Phi_\e\cdot {jU_{\e'}^\delta}.
\]
The second term on the right-hand side converges to $\int_{\R^3}  \chi\rho \Phi\cdot v_\delta
= \int_{\R^3} \rho  \Phi\cdot v_\delta$ as $\e \to 0$. The proof of this statement is like that
of \eqref{GP.conv1}, but easier.
We break the other term into two pieces.
The first is
\begin{align*}
\frac 1{\logeps^2}H_\e( \chi U^\delta_{\e'}; B_R)
&
=
\frac 1{\logeps^2}\int_{B_R}f_\e^2 \left( \frac 12 |d U^\delta_{\e'}|^2 + \frac {f_\e^2}{4\ep^2}(|U^\delta_{\e'}|^2-1)^2 \right)\\
&\le
\frac{(1+o(1))}{|\log \e'|^2}\int_{B_R}f_\e^2 \left( \frac 12 |d U^\delta_{\e'}|^2 + \frac 1{4\ep'^2}(|U^\delta_{\e'}|^2-1)^2 \right).
\end{align*}
Then, since $\supp(\rho)\subset \bar \Omega\subset\subset B_R$, it follows from \eqref{vepdelta} and the uniform convergence $f_\e^2\to \rho$ that
\[
\limsup_{\e\to 0}\frac 1{\logeps^2}H_\e( \chi U^\delta_{\e'}; B_R)\le \int \rho(|v_\delta|^2 + |dv_\delta|).
\]
And from properties of $\chi$ and exponential smallness of $f_\ep$ outside of $B_R$,
it easily follows that $H_\e( \chi U^\delta_{\e'}; \R^3\setminus B_R) \to 0$ as $\e\to 0$.
By combining the above inequalities, we find that
\[
\limsup_{\e\to 0}
\frac 1{\logeps^2}(\calG_\e(u^\delta_\e) - \calG_\e(f_\e))  \ \le \calG(v_\delta)\le \calG(v) + C \delta.
\]
Note also that
\[
\frac{ ju_\e^\delta}{\logeps} = (1+ o(1))(f_\e \, \chi)^2  \frac{jU_{\e'}^\delta}{|\log \e'|} \to \rho v_\delta\mbox{ weakly in }L^{4/3}(\R^3).
\]
Conclusion (iii) now follows by setting $u_\e := u_\e^{\delta(\e)}$ for $\delta(\e)$ converging to $0$
sufficiently slowly.
\end{proof}

We conclude this section with the proof of the approximation lemma used above.

\begin{proof}[Proof of Lemma \ref{lem:GPdense}]
We introduce some auxiliary functions.
First, for $r\in (0,1]$, let 
\[
\Omega_r := \{ x\in \R^3: \dist(x, \Omega)<r \},
\quad\quad\quad
\Omega_{-r} := \{ x\in\Omega : \dist(x, \partial \Omega) > r \}.
\]
Next, for sufficiently small $\sigma>0$, let  $\Psi_\sigma:\Omega_{\sigma}\to \Omega$ be the $W^{1,\infty}$ diffeomorphism 
given by
\[
\Psi_\sigma(x) := \begin{cases}
x -  \sqrt\sigma (d(x) +\sqrt \sigma-  \sigma) \bar\nu(x)&\mbox{ if }x\in \Omega_{\sigma} \setminus\Omega_{\sigma - \sqrt \sigma}\\
x&\mbox{ if }x\in\Omega_{\sigma - \sqrt \sigma}
\end{cases}
\]
where $\bar \nu(x)$ is the outer unit normal to $\partial \Omega$ at the point of $\partial \Omega$
closest to $x$, and $d(x)$ is the signed distance (positive outside $\Omega$, negative in $\Omega$) to $\partial \Omega$. 
Note that
\beq
\| D\Psi_\sigma - I \|_\infty \le  C \sqrt\sigma,
\quad\quad\quad
\| D\Psi_\sigma^{-1} - I  \|_\infty \le  C \sqrt \sigma
\label{DPsi.bds}\eeq
where $I$ denotes the identity matrix. In addition, $\rho(\Psi_\sigma(x)) \ge  \rho(x)$ for all $x$, whenever $\sigma$ is sufficiently small, since $D\rho(y) = - c(y)\nu(y)$ for $y\in \partial \Omega$, with $c(y) \ge c>0$ for all $y$, by \eqref{lambda.reg}.

Next, let $\chi_\sigma\in C^\infty_c(\R^3)$ be a nonnegative function such that 
$\chi_\sigma = 1$ in $\Omega_{\sigma/4}$
and $\chi_\sigma$ has compact support in $\Omega_{3\sigma/4}$.
Finally, for $\tau\in (0,1]$ let $\eta_\tau$ be a smooth nonnegative even mollifier with supported in $B(0, \tau)$ with $\int \eta_\tau = 1$.

We define $v_\delta := \eta_\tau \, * \, (\chi_\sigma \, \cdot \, \Psi_\sigma^\# v)$,
where $\tau, \sigma$ will be fixed below.
Here $\Psi_\sigma^\# v$ denotes the pullback of $v$ by $\Psi_\sigma$, which is a one-form on
$\Omega_\sigma$, and the product $\chi_\sigma \, \cdot \, \Psi_\sigma^\# v$ is understood to
equal zero on $\R^3\setminus \Omega_\sigma$. 

Note that $\chi_\sigma \cdot  \Psi_\sigma^\# v$ is integrable on $\R^3$, so that the convolution in the definition of $v_\delta$ makes sense. Indeed, \eqref{DPsi.bds} implies that
$
| \Psi_\sigma^\# v(x)| \le (1+ C\sqrt\sigma)|v(\Psi_\sigma(x))|$ for all $x$, so that by a change
of variables,
\begin{align*}
\int_{\R^3}
|\chi_\sigma \cdot  \Psi_\sigma^\# v|
\le
\int_{\Omega_{3\sigma/4}} |\Psi_\sigma^\# v|
\le
C \int_{\Psi_\sigma(\Omega_{3\sigma/4})} |v|
%&\le C \int_{\Psi_\sigma(\Omega_{3\sigma/4})} \rho |v| \ \ 
\ \ < \ \ \infty.
\end{align*}
The final estimate follows from the $v\in L^2_\rho$, as well as
the fact that  $\rho$ is bounded away from $0$ in 
$\Psi_\sigma(\Omega_{3\sigma/4})$, since this set is compactly contained in $\Omega$.

We will take $\tau < \sigma/4$, so that $v_\delta = \eta_\tau* \Psi_\sigma^\# v$ in $\Omega$.
Then
\[
\int_\Omega \rho |v_\delta - v|^2  
\le
2 \int_\Omega \rho |\eta_\tau* \Psi_\sigma^\# v - \Psi_\sigma^\# v |^2
+   
2 \int_\Omega \rho | \Psi_\sigma^\# v - v |^2.
\]
The definition of $\Psi_\sigma$ implies that
$\Psi_\sigma^\# v - v = 0$ if $\dist(x, \R^3\setminus \Omega) > \sigma$, and 
$|\Psi_\sigma^\# v - v| \le C|v|$, so that the second term on the right-hand side tends to $0$
as $\sigma\to 0$, by the dominated convergence theorem, and can be made less than $\delta/2$
by choosing $\sigma$ appropriately. Then we can clearly make the first term on the right-hand side less than $\delta/2$ by taking $\tau$ smaller if necessary.

To estimate $\int \rho|dv_\delta|$, we consider the action of $dv_\delta$ on some $\phi \in C^\infty_c(\Lambda^2\Omega)$. Exactly as in \eqref{mol.rewrite}, we can rewrite
\[
\int_\Omega \rho \, \phi \cdot d v_\delta
=
\int_{\Psi_\sigma(\Omega)} v \cdot d^* \star  (\Psi_{\sigma}^{-1})^\# (\eta_\tau* ( \rho \,\star \phi )).
\]
Since $d^* \star  (\Psi_{\sigma}^{-1})^\# (\eta_\tau* ( \rho \,\star \phi ))$ is smooth, and thus continuous,
\[
\int_\Omega \rho \, \phi \cdot d v_\delta
 \ = \ \int_{\Psi_\sigma(\Omega)} dv \cdot \star  (\Psi_{\sigma}^{-1})^\# (\eta_\tau* ( \rho \,\star \phi ))
\]
where the right-hand side indicates the integral of the continuous function $\star  (\Psi_{\sigma}^{-1})^\# ( \cdots)$
with respect to the measure $dv$.
It follows that \begin{align*}
\int_\Omega \rho \, \phi \cdot d v_\delta
&\le
\sup_{x\in \Psi_\sigma(\Omega)}\left| \frac 1\rho  (\Psi_{\sigma}^{-1})^\# (\eta_\tau* ( \rho \,\star \phi ))(x)\right| \int_\Omega \rho |dv|.
\end{align*}
And for $x\in \Psi_\sigma(\Omega)$, since $|\star \phi (x)| = |\phi(x)|$ for all $x$,
\begin{align}
\left| \frac 1\rho  (\Psi_{\sigma}^{-1})^\# (\eta_\tau* ( \rho \,\star \phi ))(x)\right| 
&\overset{\eqref{DPsi.bds}}\le 
\frac {(1+ C\sqrt\sigma)}{\rho(x)} \ (  \eta_\tau * | \rho \,\star \phi |)(\Psi_\sigma^{-1}(x) )
\nonumber\\
&\le 
\| \phi\|_{\infty} \frac {(1+ C\sqrt\sigma)}{\rho(x)} \ (  \eta_\tau *  \rho  )(\Psi_\sigma^{-1}(x) )
\nonumber\\
&\le 
\| \phi\|_{\infty} \frac   {(1+ C\sqrt\sigma)} {\rho(x)} \left( \sup_{B_\tau(\Psi_\sigma^{-1}(x) )}\rho\right).
%\\&\le 
%\| \phi\|_{\infty} \frac   {(1+ C\sqrt\sigma)} {\rho(x)} \left( \rho(\Psi_\sigma^{-1}(x) ) +C \tau |D\rho(\Psi_\sigma^{-1}(x))|\right).
\label{dv.technical}\end{align}
And writing $y := \Psi_\sigma^{-1}(x)$, if $\sigma$ is small enough then
$D\rho \cdot \nu \le - c < 0$ in the set where $\Psi_\sigma$ is not the identity, so it follows from
the mean value theorem and the definition of $\Psi_\sigma$ that
$\rho(\Psi_\sigma(y)) \ge  \rho(y) + c \sqrt\sigma(d(y)  +\sqrt\sigma-\sigma)^+$, where $(\cdots)^+$ denotes the positive part. Thus
\[
\frac 1 {\rho(\Psi_\sigma(y))} \left( \sup_{B_\tau(y )}\rho\right)\
\le
 \frac{ \rho(y) + C\tau }{ \rho(y) + c \sqrt\sigma(d(y)  +\sqrt\sigma-\sigma)^+ }.
\]
We insist that $\sigma< 1/16$ (in addition to other smallness conditions), so that
in $\Omega\setminus \Omega_{-\sigma}$, we have the inequality
$d(y)+ \sqrt\sigma - \sigma \ge \sqrt \sigma- 2\sigma \ge \frac 12 \sqrt \sigma$.
In this set, then,
\[
 \frac{ \rho(y) + C\tau }{ \rho(y) + c \sqrt\sigma(d(y)  +\sqrt\sigma-\sigma)^+ }
\le 
 \frac{ \rho(y) + C\tau }{ \rho(y) +  (c/2)\sigma}
\ \le 1
\quad\mbox{ for $\tau$ sufficiently small}.
\]
By taking $\tau$ small enough, we can make  $C\tau  /\rho(y)$ as small as we like
in the set $\Omega_{-\sigma}$, where $\rho \ge c \sigma$.
Thus by taking $\tau$ still smaller, if necessary, we can guarantee that the right-hand side of
\eqref{dv.technical} is bounded by $(1+\delta)\| \phi\|_\infty$.
Inserting this into the above estimates, we conclude that
\[
\int_\Omega \rho \, \phi \cdot d v_\delta
\le \|\phi\|_\infty (1+\delta) \int_\Omega \rho|dv|
\]
for all $\phi\in C^\infty_c(\Lambda^2\Omega)$, and hence that $v_\delta$ satisfies \eqref{GP.dense}.
\end{proof}

\subsection{a dual problem and critical forcing}

In this section we give the proof of Theorem \ref{thm:GPcritical}. We will use notation introduced in
Section \ref{sect:GPintro}.

\begin{proof}[Proof of Theorem \ref{thm:GPcritical}]
{\bf 1}. We first formulate a dual problem.
Let
\[
I_0(w) := -\int_\Omega \rho w \cdot \Phi +  \frac12 \int_\Omega \rho |dw|  \ \ = \ \ - (w,\Phi)_\rho + \frac 12  \int_\Omega \rho |dw| 
\]
so that $\calG(v) = I_0(v) +  \frac 12 \| v\|_\rho^2$.
Then 
\[%\begin{align*}
I_0^*(v) 
:=	%&:= 
\sup_{w\in L^2_\rho}\left(  (v,w)_\rho - I_0(w)\right)\\
= 	%&=
\sup_{w\in L^2_\rho}\left(  (v + \Phi,w)_\rho - \frac 12  \int_\Omega \rho |dw| 
\right).
\]
It is clear that if $ (v + \Phi,w)_\rho - \frac 12  \int_\Omega \rho |dw| >0$ for any $w$, then the supremum
on the right-hand side above is unbounded, so we conclude that
\[
I_0^*(v)
=\begin{cases}
0	&\mbox{ if } v + \Phi \in N \ \ \ \mbox{(defined below)}\\
+\infty &\mbox{ if not},
\end{cases}
\]
for 
\beq
N := \left\{ \xi\in L^2_\rho(\Lambda^1\Omega) : (\xi, w)_\rho \le  \frac 12 \int_{\Omega} \rho |dw| \ \mbox{ for all }w\in L^2_\rho(\Lambda^1\Omega) \right\}.
\label{GP:Ndef}\eeq
Then it follows from basic facts about duality, see Lemma \ref{lem:dual}, that the unique minimizer $v_0$ 
of $\calG$ in $L^2_\rho$ is also the unique minimizer of 
\beq
\calG^\dagger(v) := I_0^*(-v) + \frac 12 \|v\|_\rho^2 = 
\begin{cases}
\frac 12 \|v\|_\rho^2 &\mbox{ if }\Phi-v\in N\\
+\infty		&\mbox{ if not}.
\end{cases}
\label{GP:dual1}\eeq

{\bf 2}. We next rewrite the dual problem. It is immediate from the definition \eqref{GP:Ndef} of $N$
that $N\subset (\ker d)_\rho^\perp$. Hence,
writing
$P_\rho$ for $L^2_\rho$-orthogonal projection onto $(\ker d)_\rho$, it follows that
\[
\mbox{$\Phi-v\in N$ \ \ \ \ \ \  if and only if } \ \ \   \   \ \mbox{$P_\rho \Phi = P_\rho v$  
\ {\em  and } \  
$P_\rho^\perp\Phi  -  P_\rho^\perp v \in N$}
\]
In particular,
\[
\frac 12 \|v \|_\rho^2 = \frac 12 \|  P_\rho \Phi\|_\rho^2
+  \frac 12 \|P_\rho^\perp v\|^2\quad\mbox{ if $\Phi-v\in N$}
\]
so that minimizing the $L^2_\rho$ norm of $v$, subject to the constraint $\Phi-v\in N$, is equivalent 
to minimizing the $L^2_\rho$ norm of $P_\rho^\perp v$, subject to the constraint $P_\rho^\perp\Phi - P_\rho^\perp v\in N$.

Now recall from  the description \eqref{kerdperprho} of $(\ker d)_\rho^\perp = \operatorname{Image}(P_\rho^\perp)$
that every element of $(\ker d)_\rho^\perp$ can be written in the form $\frac {d^*\beta}\rho$ for some
$d^*\beta \in H^1_N(\Lambda^2\Omega)$. In particular, if we write $P_\rho^\perp\Phi = \frac{d^*\beta_\Phi}\rho$
and  $P_\rho^\perp v_0 = \frac{d^*\beta_0}\rho$, then
\begin{align}
&v_0 = P_\rho\Phi + \frac {d^*\beta_0}\rho\quad\mbox{ where }\beta_0\mbox{ minimizes }
\nonumber\\
&\quad\quad\beta\mapsto \frac 12 \| \frac {d^*\beta}\rho\|_\rho^2
\mbox{ in  }\left\{ \beta\in H^1_N(\Lambda^2\Omega)  \ : \  
 \frac {d^*(\beta_\Phi - \beta)}\rho \in N \right\}.
\label{v0.dual11}\end{align} 
As usual, we understand $\| \frac{d^*\beta}\rho\|_\rho $ to equal  $+\infty$ if $ \frac {d^*\beta}\rho$ does not
belong to $L^2_\rho(\Lambda^1\Omega)$.

We now rewrite the constraint by noting that for any  smooth $w\in C^\infty (\Lambda^1\bar\Omega)$
and for $\beta\in H^1_N(\Lambda^2\Omega)$ such that
$\frac {d^*\beta}\rho \in L^2_\rho(\Lambda^1\Omega)$,
\[
(\frac{d^*(\beta_\Phi - \beta)}\rho, w)_\rho \ = \ 
\int_\Omega d^*(\beta_\Phi - \beta)\cdot w \ = \ 
\int_\Omega (\beta_\Phi - \beta)\cdot d w.
\]
Here the boundary terms vanish due to the fact that $(\beta_\Phi)_N = \beta_N = 0$.
Then the definition \eqref{GP:Ndef} of $N$ and facts about density of smooth
functions established in Lemma \ref{lem:GPdense} imply that
\begin{align}
 \frac {d^*(\beta_\Phi - \beta)}\rho \in N
% &\iff
%(\frac{d^*(\beta_\Phi - \beta)}\rho, w) \le  \frac 12 \int_\Omega \rho |dw|\quad \mbox{ for all }w\in L^2_\rho(\Lambda^1\Omega)\nonumber\\
&\iff
(\frac{d^*(\beta_\Phi - \beta)}\rho, w) \le  \frac 12 \int_\Omega \rho |dw|\quad \mbox{ for all }w\in C^\infty(\Lambda^1\bar\Omega)\nonumber \\
&\iff
\int_\Omega (\beta_\Phi - \beta)\cdot dw \le  \frac 12 \int_\Omega \rho |dw| \quad\mbox{ for all }w\in C^\infty(\Lambda^1\bar
\Omega)\nonumber\\
&\iff
\| \beta_\Phi - \beta\|_{\rho*}\le \frac 12
\label{rewrite.constraint}\end{align}
where we recall the definition 
\[
\| \gamma \|_{\rho*} := \sup\{ \int_\Omega\gamma \cdot dw \ : \ \ w\in C^\infty(\Lambda^1\bar\Omega), \  
\int_\Omega \rho |dw | \le 1 \}.
\]
Now by combining \eqref{v0.dual11} and \eqref{rewrite.constraint}, we obtain the characterization of $v_0$
appearing in Theorem 3, see \eqref{GP.conmin1}, \eqref{GP.conmin2}, \eqref{GP.conmin3}.

Observe further that, by stationarity of  \eqref{G0.def} with respect to variations $t\mapsto e^tv_0$ around $t=0$ we obtain
\beq\label{eq:saturate1}
\frac{1}{2}\int_\Omega\rho|dv_0|+\int_\Omega\rho v_0\cdot(v_0-\Phi)=0\, .
\eeq
Recalling that $\Phi=P_\rho\Phi+\frac{d^*\beta_\Phi}{\rho}$, we have $\rho(v_0-\Phi)=d^*\beta_0-d^*\beta_\Phi$. Inserting in \eqref{eq:saturate1} yields \eqref{eq:saturate} after integration by parts.

{\bf 3}. 
It remains to check that $dv_0=0$ if and only if $\|\beta_\Phi\|_{\rho*}\le \frac 12$.

The global minimizer of the functional $\beta\mapsto \frac 12 \| \frac{ d^*\beta} \rho\|_{\rho}^2$ in $H^1_N(\Lambda^2\Omega)$ is attained by $\beta = 0$, and this satisfies the constraint \eqref{rewrite.constraint} if and only if $\|\beta_\Phi\|_{\rho*}\le \frac 12$.

Thus  if $\|\beta_\Phi\|_{\rho*}\le \frac 12$, then $v_0 = P_\rho\Phi\in (\ker d)_\rho$, and in this case clearly $dv_0=0$.

On the other hand, if $\|\beta_\Phi\|_{\rho*} > \frac 12$, then  $v_0 - P_\rho\Phi = \frac {d^*\beta_0}\rho$ is a (nonzero) element of
$(\ker d)_\rho^\perp$, and hence in this case $0 \ne d(v_0 - P_\rho\Phi) = dv_0$. 

\end{proof}

\section{further remarks}\label{S:other}

\subsection{symmetry reduction}

In the presence of rotational symmetry, the functionals we study in this paper reduce to
simpler $2$-dimensional models. We discuss this first for the functional $\calG$, defined in \eqref{G0.def}, arising in case of Bose-Einstein condensates.

\begin{lemma}
Consider the functional 
$\calG(v) = \int_{\Omega} \rho \left( \frac{|v|^2}2  - v \cdot \Phi  +\frac{1}{2} |d v|\right)$,
and assume that there exist some $\tilde \Omega\subset [0,\infty)\times \R$ and
some $\tilde\rho : \tilde \Omega\to (0,\infty)$ such that
\[
\Omega = \{ (r\cos\alpha, r\sin\alpha, z) : (r,z)\in \tilde \Omega, \alpha\in \R\},
\]
\[
\rho(r\cos\alpha, r\sin\alpha, z)  = \tilde \rho(r,z)\ \ \forall \alpha\in \R.
\]
Assume moreover that there exists some $\phi:\tilde\Omega\to \R$ such that
\[
\Phi(r\cos\alpha, r\sin\alpha, z) = \phi(r,z) d\theta\quad\quad \mbox{ for all }\alpha.
\]
Then the unique minimizer $v_0$ of $\calG$ is given in cylindrical coordinates
by
$v_0 = w_0(r,z) d\theta$, where $w_0$ minimizes the functional
\beq
\calG^{red}(w) := \frac 12 \int_{\tilde\Omega} \tilde\rho\left( |\nabla w| +  \frac {(w-\phi)^2}r  \right) \, dr \, dz
\label{Greduced}\eeq
in the space of functions $w:\tilde\Omega\to \R$ such that 
$\int_{\tilde \Omega}\frac{ \tilde \rho}r \,  w^2  \, dr \, dz <\infty$.
\label{lem:reduce1}
\end{lemma}

We set $\calG^{red}(w)  = +\infty$ if $dw$ is not a Radon measure in $\tilde \Omega$ or
if $r \tilde \rho$ is not $|dw|$-integrable.

As noted in the introduction, $\calG^{red}$ is exactly a (weighted) version of a
functional that has been studied in the context of image denoising, 
see for example \cite{ROF, CasChamNov}.

\begin{proof}
{\bf 1}. Let $R_\alpha:\R^3\to \R^3$ denote rotation by an angle $\alpha$ around the $x_3$ axis.
%, so that
%\[
%R_\alpha 
%\left(\begin{array}{c}x_1\\x_2\\x_3
%\end{array}\right)
% := \left(\begin{array}{ccc}
%\cos \alpha &\sin \alpha &0\\
%-\sin\alpha&\cos\alpha &0\\
%0&0&1\end{array} \right)
%\left(\begin{array}{c}x_1\\x_2\\x_3
%\end{array}\right).
%\]
Equivalently, in cylindrical coordinates, $R_\alpha$ is the map
$(r,\theta, z)\mapsto (r,\theta+\alpha, z)$.
Then our assumptions imply that 
\[
R_\alpha(\Omega) = \Omega, 
\quad\quad
\rho\circ R_\alpha = \rho.
\quad\quad
R_\alpha^\#\Phi = \Phi
\]
for all $\alpha$. It easily follows that $\calG(R_\alpha^\# v) = \calG(v)$ for all $v$ and $\alpha$.
By uniqueness of the minimizer $v_0$ of $\calG$, which follows from strict convexity, we conclude that 
\[
v_0\in \tilde L^2_\rho(\Lambda^1\Omega) :=  \left\{ v\in L^2_\rho(\Lambda^1\Omega) \  : \ 
R_\alpha^\# v = v\mbox{ for all }\alpha \right\}.
\]
It is then immediate that $v_0$ minimizes $\calG$ in $ \tilde L^2_\rho(\Lambda^1\Omega)$.

{\bf 2}. Any $v\in \tilde L^2_\rho(\Lambda^1\Omega)$ can be written in polar coordinates as
\beq
v = v^\theta(r,z) d\theta + v^r(r,z) dr + v^z(r,z) dz.
\label{eq:symforms}\eeq
We claim that for any such $v$, 
%%% start replacement %%%%
\beq
\calG(v) \ge \calG(v^\theta d\theta) = 2\pi \calG^{red}(v^\theta) - C(\Phi).
\label{nearlyobvious}\eeq
Clearly, this together with Step 1 implies the conclusion of the lemma.
To prove \eqref{nearlyobvious}, note that if $v$ is smooth and has the form
\eqref{eq:symforms}, then
\[
|dv| = | \partial_r v^\theta dr\wedge d\theta + \partial_z v^\theta dz\wedge d\theta
+ (\partial_r v^z - \partial_z v^r) dr\wedge dz| \ge |d(v^\theta d\theta)|.
\]
In the general case,
the same conclusion follows from the density of smooth 
functions in $v\in L^2_\rho(\Lambda^1\Omega)$ in the sense of 
Lemma \ref{lem:GPdense}.
It is also clear that if $v\in \tilde L^2_\rho(\Lambda^1\Omega)$ and $\Phi = \phi(r,z)d\theta$, then
\[
|v-\Phi|^2 =  \frac{(v^\theta - \phi)^2}{r^2} + (v^r)^2 + (v^z)^2 \ge \frac{(v^\theta - \phi)^2}{r^2}
= |v^\theta d\theta - \Phi|^2
\]
pointwise, so that
\[
\int_\Omega \rho (\frac{|v|^2}2 - v \cdot\Phi) 
=
\int_\Omega \frac  \rho 2\left(|v-\Phi|^2 - |\Phi|^2\right)
\ge
\int_\Omega \rho (\frac{|v^\theta d\theta |^2}2 - (v^\theta d\theta) \cdot\Phi) .
\]
Combining these estimates, we conclude that $\calG(v) \ge \calG(v^\theta d\theta)$ for all $v\in \tilde L^2_\rho(\Lambda^1\Omega)$.

Finally, the identity
\[
\calG(v^\theta d\theta) =   2\pi \, \calG^{red}(v^\theta) - C(\Phi), \quad\quad\quad
C(\Phi) := \int_\Omega \frac{\rho}2 |\Phi|^2 % = \pi \int_{\tilde \Omega}\tilde \rho \frac{\phi^2}r \ dr \ dz 
\]
is clear if $v^\theta$ is smooth, and in the general case can be verified either by a density argument similar to the one given above, or
by directly relating the definitions of the total variation measures associated with $d(v^\theta d\theta)$ in $\Omega$ and $\nabla v^\theta$ in $\tilde \Omega$, respectively.
%%% start older %%%%

\end{proof}

\begin{remark}As in Theorem  \ref{thm:GPcritical}, one can 
use duality to rewrite the problem of minimizing 
$\calG^{red}$ as a constrained variational problem. 
For example, one can verify that $v_0$ minimizes $\calG^{red}$
if and only if it minimizes the functional
\beq
w\mapsto \int_{\tilde \Omega} \frac{\tilde \rho}r w^2 \ dr \, dz
\label{reducedL2}\eeq
subject to the constraint
\beq
\int_{\tilde \Omega}  \frac{\tilde \rho}r  (\phi-w) \zeta \ dr \ dz \le \frac 12 \int_{\tilde \Omega} \tilde \rho |\nabla \zeta|
\quad\quad\quad\mbox{for all $\zeta \in C^\infty(\tilde \Omega)$},
\label{reducedconstraint}\eeq
analogous to \eqref{GP:dual1}.
One could also 
reformulate this as a problem
of minimizing  a weighted Dirichlet energy of
a $1$-form on $\tilde \Omega$ with a nonlocal constraint
like that of  \eqref{GP.conmin2}, but in this setting
this seems to us less natural, since the formulation in terms of functions rather
than $1$-forms seems simpler.
\label{rem:reduce}\end{remark}

\begin{remark}
For velocity field represented by the $1$-form $v = w(r,z)d\theta$,  the associated
vorticity $2$-form is $dv = \partial_r w\,  dr \wedge d\theta + \partial_z w\,  dz\wedge d\theta$.
The vorticity vector field, that is, the vector field  dual to $dv$, is then $\frac 1r (\partial_r w \, \hat e_z -  \partial_z w\, \hat e_r)$,
where $\hat e_z$ and $\hat e_r$ denote unit vectors in the (upward) vertical and (outward) radial directions  respectively. 
It is natural to interpret integral curves of this vector field as ``vortex curves''.
Since the vorticity vector field has no $\hat e_\theta$ component and is always
tangent to level surfaces of $w$, we conclude that, formally, vortex curves have the form
``$\theta  = \mbox{constant}, w =  \mbox{constant}$" (at least for regular values of $w$).

Thus in the reduced 2d model, we interpret level sets of a minimizer $w_0$, or
more precisely sets of the form $\partial \{ (r,z) : w_0(r,z)>t\}$, as representing vortex curves.

For similar reasons, one should think of the ``vorticity measure " as being given by
$\nabla^\perp w_0$, rather than $\nabla w_0$. 
\label{reducedvc}
\end{remark}

Similarly, we have

\begin{lemma}
Consider the functional 
\[
\calF(v,A) = \frac 12 \int_{\Omega} |d v| + |v-A|^2 + \frac 12 \int_{\R^3}|d(A-A_{ex})|^2 
\]
and assume that there exist some $\tilde \Omega\subset [0,\infty)\times \R$ and
$\phi:\tilde\Omega\to \R$ such that
\[
\Omega = \{ (r\cos\alpha, r\sin\alpha, z) : (r,z)\in \tilde \Omega, \alpha\in \R\},
\]
\[
A_{ex}(r\cos\alpha, r\sin\alpha, z) = \phi(r,z) d\theta\quad\quad \mbox{ for all }\alpha.
\]
Then the unique minimizer $(v_0, A_0)$ of $\calF$ is given in cylindrical coordinates
by
$(v_0, A_0) = (w_0(r,z) d\theta, b_0(r,z)d\theta) $, where $(w_0, b_0)$ minimizes the functional
\beq
\calF^{red}(w,b) := \frac 12 \int_{\tilde\Omega}  |\nabla w| +  \frac {(w- b)^2}r \, dr\, dz \ + \  
\frac 12\int_{\tilde \R^2} \frac{|\nabla(b-\phi)|^2}r   \, dr \, dz
\label{Freduced}\eeq
in the space of $(w,b):\tilde \R^2\to \R^2$ for which $\calF^{red}(w,b)$ is well-defined and finite,
where $\tilde \R^2 := \{(r,z) : r>0\}$.
\label{lem:reduce2}\end{lemma}

We omit the proof, which is extremely similar to that of Lemma \ref{lem:reduce1}. 

\subsection{contact curves and vortex curves}
It is interesting to ask whether one can define a useful analog of the
``contact set" (as normally defined for classical obstacle problems) for the variational problems with nonlocal constraints formulated
in Theorems \ref{thm:GLmain} and \ref{thm:GPcritical}.
We address this question first for Bose-Einstein condensates in
the presence of rotational symmetry, as discussed immediately above.
Thus, we assume that $w_0:\tilde \Omega\to \R$ minimizes the functional
\eqref{reducedL2} subject to the constraint \eqref{reducedconstraint}.
An approximation argument starting from  \eqref{reducedconstraint} shows that
if $E$ is a set of locally finite perimeter in $\tilde \Omega$, then
\beq
\int  \frac {\tilde \rho}r (\phi-w_0) \chara_E \ dr \ dz  \le \frac 12 \int \tilde \rho |\nabla \chara_E|.
\label{eq:sept24}\eeq
We say that $\partial E$ is a {\em contact curve} if equality holds in the above (where $\partial E$ should be understood as the $1$-dimensional set that carries $|\nabla \chara_E|$).

\begin{lemma}
For a.e. $t$, $\partial \{ w_0> t\}$ is a contact curve.
\label{lem:curves1}\end{lemma}

As argued in Remark \ref{reducedvc}, it is natural to interpret $\partial \{ w_0> t\}$ 
as a ``vortex curve'',
so the Lemma states, heuristically, that every vortex curve for $w_0$ is also a contact curve.

\begin{proof}By using rotational symmetry to reduce \eqref{eq:saturate} to the $(r,z)$ variables,
or by using the fact that $0 = \left. \frac d{dt} \calG^{red}(e^tw_0)\right|_{t=0}$, we find that
\[
\frac 12 \int \tilde \rho |\nabla w_0|
+\int  \frac {\tilde \rho}r (w_0 - \phi) w_0 \ dr \ dz 
= 0.
\]
Using the coarea formula, we rewrite this as
\beq
\int_{-\infty}^\infty \left( \frac 12 \int \tilde \rho |\nabla \chara_{\{ w_0>t\}}| \ + 
\int \frac {\tilde \rho}r (w_0-\phi)  \chara_{\{ w_0>t\}} \, dr \, dz \right) dt = 0.
\label{eq:ccurvevcurve}\eeq
It follows from \eqref{eq:sept24} that  
\[
\frac 12 \int \tilde \rho |\nabla \chara_{\{ w_0>t\}}| \ + 
\int \frac {\tilde \rho}r (w_0-\phi)  \chara_{\{ w_0>t\}} \, dr \, dz \ge 0
\]
for every $t$, and then \eqref{eq:ccurvevcurve}
implies that in fact equality holds for a.e. $t$.
\end{proof}

It is almost certainly not true that every contact curve for the minimizer
$w_0$ is also a vortex curve, in the generality that we consider here,
due to the possibility of degenerate (nonlocal) obstacles, as in the classical obstacle
problem.  One might hope, however, that the vortex curves and contact curves
coincide under reasonable physical 
assumptions (for example, $\Phi = r^2d\theta$, corresponding to rotation of a condensate around the $z$ axis, probably also with some conditions on $\rho$.)

\medskip

The situation is more complicated for Bose-Einstein condensates in a general domain $\Omega\subset\R^3$ without rotational symmetry, since in this case the
analogs of vortex curves and contact curves may
not in fact be curves and do not in general
admit a very easy concrete characterization.
Abstractly, they may be described as follows:
%Closely related questions are studied  in \cite{smirnov}, which characterizes decompositions of  divergence-free vector-valued measures fields on $\R^n$. 
if we write $\mathcal Z$ to denote the closure (in the sense of distributions)
of   
\[
\{ d\alpha : \alpha\in L^2(\Lambda^1\Omega), \int_\Omega \rho |d\alpha| \le 1\},
\]
then one can think of the set $\extr \mathcal Z$ of extreme points of (the convex set) $\mathcal Z$
as analogous to the objects --- distributional boundaries of sets of finite weighted perimeter --- used above
to describe vortex and contact curves.
Indeed, by arguments  exactly like those of Remark 3 of \cite{smirnov}, general convexity considerations and a bit of functional analysis imply that
$\extr \mathcal Z$ is a nonempty Borel subset of a metric space, and for any $T$ in the vector space generated by $\mathcal Z$ (that is, the space $\cup_{\lambda >0}\lambda \mathcal Z$),
there is a measure $\mu_T$ on $\extr\mathcal Z$ such that
\beq
T  =  \int_{\extr\mathcal Z} \omega \ d\mu_T(\omega)
\label{extrZ1}\eeq
and in addition
\beq
\int_\Omega \rho \,d|T| \ = \ \int_{\extr\mathcal Z}\left( \int_\Omega \rho \,d| \omega| \right) \ d\mu_T(\omega) .
\label{extrZ2}\eeq
We remark that in the closely related situation of divergence-free vector fields on $\R^n$, a concrete characterization of elements of the analog of $\extr\mathcal Z$ as ``elementary solenoids"  is established in \cite{smirnov}.

With this notation, an analog of Lemma \ref{lem:curves1} is

\begin{lemma}
Let $\beta_0$ be the minimizer of the constrained variational problem
\eqref{GP.conmin1}, \eqref{GP.conmin2}), so that
$v_0 = P_\rho \Phi + \frac {d^*\beta_0}\rho$ is the minimizer of $\calG(\cdot)$.
Then
\beq
\int_{\Omega} (\beta_\Phi - \beta_0) \cdot d\omega \ \le \ \frac 12 \int_\Omega \rho d|\omega|.
\label{const3}\eeq
for every $\omega \in \mathcal Z$. We say that $\omega\in \mathcal\extr Z$ is a `` generalized contact curve"
if the above holds with equality.

Furthermore, let $\mu_{dv_0}$ denote a measure on $\extr\mathcal Z$ 
satisfying \eqref{extrZ1}, \eqref{extrZ2} (with $T$ replaced by $dv_0$). Then $\mu_{dv_0}$ a.e. 
$\omega$ is a generalized contact curve.
\label{lem:curves2}\end{lemma}

The proof is exactly like that of Lemma \ref{lem:curves1}, except that \eqref{extrZ1}, \eqref{extrZ2} are substituted for the coarea formula. Then \eqref{const3} follows immediately from the fact that $\beta_0$ satisfies \eqref{GP.conmin2}, and the last assertion us a consequence of 
\eqref{eq:saturate}.

A version of Lemma \ref{lem:curves2} could be formulated for the functional
$\calF$ arising in the description of superconductivity and the associated contained
variational problem described in Theorem \ref{thm:GLmain},  using a measurable
decomposition \eqref{extrZ1}, \eqref{extrZ2} of the vorticity $dv_0$ to deduce from \eqref{eq:B0dv0} a
precise form of the assertion that every (generalized) vortex curve is a (generalized) contact curve.
%We state without proof 
%the parallel result  for minimizers of $\calF^{red}$: 

It would presumably be rather easy to adapt results of \cite{smirnov} to the closely related situations
considered here, to obtain concrete descriptions of $\extr \mathcal Z$, or the corresponding objects relevant for superconductivity, although we are not sure that this would add much insight. It would also be interesting to know whether, if we consider the model case
of uniform rotation about the $z$ axis (for Bose-Einstein) or a constant applied magnetic field (for Ginzburg-Landau), the complexities sketched above do not in fact occur, and the vortex curves and contact curves for minimizers can in fact be identified with curves of finite length; this seems likely to us to be the case.

%\begin{lemma}
%...
%\end{lemma}

\bibliographystyle{plain}

\bibliography{mybibliography}

\end{document}